\documentclass[DIV=12,numbers=noenddot]{scrartcl}

\usepackage[utf8]{inputenc}
\usepackage[T1]{fontenc}
\usepackage{lmodern}
\usepackage{amsmath,amssymb}
\usepackage[usenames]{xcolor}
\usepackage{scalefnt}
\usepackage{xspace}
\usepackage{listings}
\usepackage{booktabs}
\usepackage{authblk}
\usepackage[numbers,sort&compress]{natbib}
\usepackage[font=small,labelfont=bf,format=plain,margin=0.05\textwidth]{caption}
\usepackage{subcaption}

\newcommand{\mytitle}{Precise calculation of the W boson pole mass beyond the Standard Model with FlexibleSUSY}
\newcommand{\myauthors}{Peter Athron, Markus Bach, Douglas H.J. Jacob, Wojciech Kotlarski, Dominik Stöckinger, Alexander Voigt}

\usepackage[
  pdftitle={\mytitle},
  pdfauthor={\myauthors},
  pdfkeywords={FlexibleSUSY,W,boson,mass},
  bookmarks=true, linktocpage, colorlinks=true, allbordercolors=white,
  allcolors=blue]{hyperref}
\usepackage{orcidlink} 

\title{\mytitle}
\date{}
\author[1]{Peter Athron\orcidlink{0000-0003-2966-5914}}
\author[2]{Markus Bach}
\author[3]{Douglas H. J. Jacob\orcidlink{0000-0002-8950-2853}}
\author[4]{Wojciech Kotlarski\orcidlink{0000-0002-1191-6343}}
\author[2]{Dominik St\"{o}ckinger}
\author[5]{Alexander Voigt\orcidlink{0000-0001-8963-6512}}
\affil[1]{Department  of  Physics  and  Institute  of  Theoretical  Physics,  Nanjing  Normal  University, Nanjing, Jiangsu 210023, China}
\affil[2]{Institut f\"ur Kern- und Teilchenphysik, TU Dresden, Zellescher Weg 19, 01069 Dresden, Germany}
\affil[3]{School of Physics and Astronomy, Monash University, Melbourne, Victoria 3800, Australia}
\affil[4]{National Centre for Nuclear Research, Pasteura 7, 02-093 Warsaw, Poland}
\affil[5]{Fachbereich Energie und Biotechnologie, Hochschule Flensburg, Kanzleistraße 91--93, 24943 Flensburg, Germany}

\newcommand{\abbrev}[1]{{\scalefont{.9}\text{#1}}} 
\newcommand{\FS}{\texttt{Flex\-ib\-le\-SUSY}\@\xspace}
\newcommand{\Himalaya}{\texttt{Hi\-ma\-la\-ya}\@\xspace}
\newcommand{\FeynHiggs}{\texttt{Feyn\-Higgs}\@\xspace}
\newcommand{\SM}{\ensuremath{\abbrev{SM}}\xspace}
\newcommand{\BSM}{\ensuremath{\abbrev{BSM}}\xspace}
\newcommand{\SSM}{\ensuremath{\abbrev{SSM}}\xspace}
\newcommand{\MRSSM}{\ensuremath{\abbrev{MRSSM}}\xspace}
\newcommand{\MSSM}{\ensuremath{\abbrev{MSSM}}\xspace}
\newcommand{\THDM}{\ensuremath{\abbrev{2HDM}}\xspace}
\newcommand{\NMSSM}{\ensuremath{\abbrev{NMSSM}}\xspace}
\newcommand{\CDF}{\ensuremath{\abbrev{CDF}}\xspace}
\newcommand{\MSbar}{\ensuremath{\overline{\abbrev{MS}}}\xspace}
\newcommand{\DRbar}{\ensuremath{\overline{\abbrev{DR}}}\xspace}
\newcommand{\OS}{\ensuremath{\abbrev{OS}}\xspace}
\newcommand{\tree}{\ensuremath{\text{tree}}\xspace}
\newcommand{\aem}{\ensuremath{\alpha_{\text{em}}}\xspace}
\newcommand{\as}{\ensuremath{\alpha_s}\xspace}
\def\imath{\ensuremath{\textrm{i}}}

\newcommand{\secref}[1]{Section~\ref{#1}}
\newcommand{\figref}[1]{\figurename~\ref{#1}}
\newcommand{\tabref}[1]{\tablename~\ref{#1}}
\newcommand{\code}[1]{\lstinline|#1|} 

\newcommand{\unity}{\mathbf{1}} 
\newcommand{\TeV}{\,\text{TeV}}
\newcommand{\GeV}{\,\text{GeV}}
\newcommand{\MeV}{\,\text{MeV}}
\newcommand{\MSUSY}{\ensuremath{M_\abbrev{SUSY}}\xspace}

\newcommand{\sarah}{\texttt{SARAH}\@\xspace}
\newcommand{\spheno}{\texttt{SPheno}\@\xspace}

\begin{document}
\maketitle
\begin{abstract}
  We present an updated calculation of the $W$ boson pole mass in
  models beyond the Standard Model with \FS.  The calculation has a
  decoupling behavior and allows for a precise $W$ pole mass
  prediction up to large new physics scales.  We apply the calculation
  to several Standard Model extensions, including the \MRSSM\ where we
  show that it can be compatible with large corrections to the $W$ boson
  mass that would be needed to fit the recent 2022 \CDF\ measurement.
\end{abstract}

\tableofcontents

\clearpage

\section{Introduction}

The pole mass of the $W$ boson, $M_W$, is a very important precision
test of the Standard Model (\SM). The relationship between the muon
decay and electroweak gauge boson masses and couplings leads to a
prediction of $M_W$ from the more precisely known values of the
muon decay constant $G_F$, the $Z$ boson pole mass $M_Z$ and the
electromagnetic coupling \aem.  Comparing this $M_W$ prediction
against the experimental measurement is a precision test of the
\SM. New physics can alter the prediction via quantum
corrections from new particles or in some cases through a modified
tree-level relationship between neutral and charged weak currents
(i.e.\ tree level corrections to the so-called $\rho$ parameter).

Updating the state-of-the-art \SM\ calculation\footnote{%
Both kinds of state-of-the-art \SM\ calculations of $M_W$ are
organized as
calculations of relationships between $M_W$ and other precisely known
electroweak quantities, which are numerically solved for $M_W$. The
\OS\ calculation is based on a relationship between $M_W$ and $M_Z$,
$G_F$ and the fine structure constant in the Thomson limit $\aem$ via
the quantity $\Delta r$. The
\MSbar\ calculation (which might also be called a mixed
\MSbar/\OS\ calculation) is based on a relationship between $M_W$,
$G_F$, the running $\aem$ and the running weak mixing angle
$\sin\theta_W$ via the quantity $\Delta\hat{r}_W$, where the running
couplings are related to their respective \OS\ counterparts via
quantities $\Delta\aem$ and $\Delta\hat\rho$.\label{foo:MSOS}}
in the on-shell (\OS)
scheme \cite{Awramik:2003rn} with the latest data\footnote{We update
  the fit formulae from Ref.\ \cite{Awramik:2003rn} with the data
  shown in \secref{sec:applications}.} we obtain
$M^{\SM,\OS}_W=80.355\GeV$ with the largest uncertainty being around
$6\MeV$ (see \secref{sec:calculation_uncertainties} for more details).
The state-of-the-art \SM\ modified minimal subtraction (\MSbar) calculation \cite{Degrassi:2014sxa}
performed at the same orders in perturbative couplings gives
$M^{\SM,\MSbar}_W=80.351\GeV$, which is $4\MeV$ lower than the
\OS\ calculation. On-shell and \MSbar\ estimates of the uncertainties
were $4\MeV$ in Ref.\ \cite{Awramik:2003rn} and $3\MeV$ in
Ref.\ \cite{Degrassi:2014sxa}, respectively, slightly lower than the
renormalization scheme dependence difference, while
Ref.\ \cite{Degrassi:2014sxa} argues that the uncertainty of the
$\MSbar$ calculation should be smaller than the difference from the
renormalization scheme dependence.  However both predictions suffer
from significant parametric uncertainties, the largest coming from the
top quark mass.  The value of $M_W$ can also be predicted through
global fits that do not include the direct measurements. Recent
results give similar predictions to the \OS\ calculation, see
e.g.\ $M^\SM_W = (80.354 \pm 0.007)\GeV$ \cite{Haller:2018nnx} and
$M^\SM_W = (80.3591 \pm 0.0052)\GeV$ \cite{deBlas:2021wap}, where in
these cases the parametric uncertainties are correctly combined in the
fit.

There are many important precision measurements of the $W$ mass that
can influence the world average.  Precision measurements of the $W$
boson mass were performed by LEP experiments, ALEPH
\cite{ALEPH:2006cdc}, DELPHI \cite{DELPHI:2008avl}, L3
\cite{L3:2005fft}, OPAL \cite{OPAL:2005rdt} and these measurements
have been taken into account in the LEP combination
\cite{ALEPH:2013dgf}.  While $W$ mass measurements are more
challenging at hadron colliders, the 2012 D0 \cite{D0:2012kms} and
2012 \CDF\ \cite{CDF:2012gpf} were able to improve the precision from
an uncertainty of $33\MeV$ from the LEP combination to a 2013 Tevatron
combination \cite{CDF:2013dpa} that has an uncertainty of $16\MeV$.
Finally a 2017 ATLAS \cite{ATLAS:2017rzl}, and a recent 2021 LHCb
\cite{LHCb:2021bjt} measurement have also been made.  The most precise
individual measurements are the ATLAS and \CDF\ measurements
($19\MeV$), while a 2021 world average of $M_W^{\text{2021}} = (80.379
\pm 0.012)\GeV$ was presented in the 2021 update to
Ref.\ \cite{ParticleDataGroup:2020ssz}.  Most recently, however,
\CDF\ were able to use their full dataset (increasing from
$2.2$~fb$^{-1}$ to $8.8$~fb$^{-1}$ of data) to significantly reduce
the uncertainty with a result of $M_W^\CDF = (80.4335 \pm 0.0094)\GeV$
\cite{shortauthordoi:10.1126/science.abk1781}.  The uncertainty is now
of a similar size to the theory uncertainty, but the central value has
shifted so that it is now $7\sigma$ away from the \SM\ prediction.
This however means that the new result is also in significant tension
with previous measurements and the 2021 world average. Nonetheless we
anticipate the new \CDF\ measurement will only increase interest in
precision predictions of $M_W$.

Indeed very recently Ref.\ \cite{Lu:2022bgw} demonstrated that new
\CDF\ $M_W$ result can be explained in the two-Higgs doublet model. In
Ref.\ \cite{Fan:2022dck}, the inert \THDM was found to be able to
explain the 2022 \CDF\ experimental result with a dark matter
candidate with a mass between $54\GeV$ and $74\GeV$.
Ref.\ \cite{Athron:2022qpo} has shown that it is possible to explain
the muon $g-2$ anomaly \cite{Muong-2:2021ojo} and the 2022
\CDF\ measurement using a pair of scalar leptoquarks mixed together.
Ref.\ \cite{Yuan:2022cpw} investigates the possibility of producing
contributions which explain the 2022 value of $M_W$ using axion-like
particle, a dark photon, or chameleon dark energy.  While they found
that an axion-like particle and the dark photon had viable parameter
regions which could explain the \CDF\ value, chameleon dark energy was
shown to be heavily constrained.  Ref.\ \cite{Strumia:2022qkt} also
found that it was possible to explain the \CDF\ value of $M_W$ through
coupling a $Z'$ boson directly to the Higgs.  Ref.~\cite{Yang:2022gvz}
finds low energy \MSSM\ scenarios that can fit the large \CDF\ $M_W$,
while evading collider limits while Ref.\ \cite{Du:2022pbp} looks at
SUSY explanations in the framework of extraordinary gauge
mediation. Ref.\ \cite{Zhu:2022tpr} looked at explanations of this
$M_W$ anomaly and the GeV antiproton/gamma-ray excesses.  The global
electroweak fits were also updated in
Refs.\ \cite{Lu:2022bgw,deBlas:2022hdk,Strumia:2022qkt}.

Here we present a new generic precision calculation of the $W$ boson pole
mass that can be used in almost any Standard Model extension via \FS
\cite{Athron:2014yba,Athron:2017fvs,Athron:2021kve}. Previously in
\FS, a \MSbar/\DRbar\ calculation was implemented
\cite{Athron:2017fvs}, which included full one-loop and leading
\SM-like two-loop contributions. However the direct
\MSbar/\DRbar\ calculation in \BSM\ models can suffer from
non-decoupling logarithms (see e.g.\ \figurename~9 of
Ref.\ \cite{Diessner:2019ebm}) generated from spurious incomplete
higher-order corrections that are included.  This can severely spoil
the precision, rendering it ill equipped for resolving between the
\SM\ and new physics models, potentially even for a very large
deviation, like the 2022 \CDF\ measurement implies.  The new calculation
resolves this problem by implementing the calculation with a strict
separation between the \SM\ contributions and the
\BSM\ contributions. The \SM\ calculation is performed at
state-of-the-art precision, while the \BSM\ part is a strict one-loop
calculation with no spurious higher-order corrections inadvertently
incorporated.

We apply this new calculation to several extensions of the Standard
Model, namely the Minimal Supersymmetric Standard Model (\MSSM), 
the Scalar Singlet Model (\SSM) and the Minimal $R$-symmetric
Supersymmetric Standard Model (\MRSSM) \cite{Kribs:2007ac}.
We demonstrate
the decoupling property of the models, and show scenarios where the
models can predict the measured values of $M_W$, including the most
recent measurement from \CDF\ in the case of the \MRSSM. In particular
this means we provide a precise interpretation of the 2022 \CDF\
measurement in the \MRSSM, demonstrating that this very well motivated
model can explain this dramatic new experimental result.

While the precision of our \BSM\ calculation is not as high as the
dedicated calculations in the
\SM\ \cite{Awramik:2003rn,Degrassi:2014sxa} and the \MSSM\
\cite{Heinemeyer:2013dia,Bagnaschi:2022qhb}, it provides the most
precise calculation of $M_W$ in many \SM\ extensions where no two-loop
calculations have been performed and matches the \SM\ precision in the
decoupling limit.  As such we believe the calculation that is
distributed with \FS\ can be a very useful tool for studying the
implications of the \CDF\ measurement.

 In the \SM\ the precision of the state-of-the-art predictions in the \OS\
 and \MSbar schemes has been improved through many contributions that have been
 carefully evaluated over the years.  They include full one-loop
 \cite{Sirlin:1980nh,Marciano:1980pb} and full two-loop
 \cite{Sirlin:1983ys,Djouadi:1987gn,Djouadi:1987di,Kniehl:1989yc,Consoli:1989fg,Halzen:1990je,Kniehl:1991gu,Barbieri:1992nz,Djouadi:1993ss,Fleischer:1993ub,Degrassi:1996mg,Degrassi:1996ps,Freitas:2000gg,Freitas:2002ja,Awramik:2002wn,Awramik:2003ee,Onishchenko:2002ve,Awramik:2002vu}
 corrections, as well as further important higher order corrections
 \cite{Avdeev:1994db,Chetyrkin:1995ix,Chetyrkin:1995js,Chetyrkin:1996cf,Faisst:2003px,vanderBij:2000cg,Boughezal:2004ef,Boughezal:2006xk,Chetyrkin:2006bj,Schroder:2005db}. It
 is challenging to apply the same level of precision in extensions of
 the Standard Model but in this work we include precisely the same \SM\
 corrections as in the state-of-the-art \SM\ \MSbar\ calculation of
 Ref.\ \cite{Degrassi:2014sxa}.

 The \MSSM\ is the most widely studied \SM\ extension and $M_W$
 calculations have been performed at the full one-loop
 \cite{Garcia:1993sb,Chankowski:1993eu,Pierce:1996zz}, building on
 earlier work in
 Refs.\ \cite{Barbieri:1983wy,Lim:1983re,Eliasson:1984yu,Hioki:1985wz,Grifols:1984xs,Barbieri:1989dc,Gosdzinsky:1990sk,Drees:1990dx,Drees:1991zk},
 and two-loop
 \cite{Djouadi:1996pa,Djouadi:1998sq,Heinemeyer:2002jq,Heinemeyer:2004gx,Heinemeyer:2006px,Heinemeyer:2013dia}
 level.  The $M_W$ calculation was also performed in the \THDM\ in Ref.\ \cite{Lopez-Val:2012uou}, 
 who found an improvement compared to the \SM\ prediction, similar to the \MSSM.  

 A closely related calculation of the $\rho$ parameter in the \THDM\ was presented in Ref.~\cite{Hessenberger:2016atw}.
 The calculation of $M_W$ has been performed in the Scalar Singlet Model in
 Ref.\ \cite{Lopez-Val:2014jva} in the on-shell scheme at the one-loop
 level. In the \NMSSM\ an early calculation was performed in
 Ref.\ \cite{Domingo:2011uf}, while Ref.\ \cite{Allanach:2013kza}
 generalised the \DRbar \MSSM\ calculation to the
 \NMSSM\ and a complete one-loop
 on-shell calculation has been performed for the \BSM\ contributions
 \cite{Stal:2015zca} with state-of-the-art \SM\ corrections and
 additional higher order SUSY corrections taken in account.  The
 \MRSSM\ predictions for $M_W$ have been performed at the one-loop
 level \cite{Diessner:2014ksa} in the \DRbar scheme and in a recent 
 on-shell calculation at the full one-loop level, which also takes
 into account all known \SM\ contributions from higher orders
 \cite{Diessner:2019ebm}.

  However, dedicated calculations in many other
  models do not exist at all, and given the number of possible
  extensions of the \SM\ it is hardly possible to carry out dedicated
  calculations that are rigorous and precise for each one. Simplistic
  estimations carried out for wider phenomenological work or rapid
  responses to new data run the risk of lacking sufficient precision
  to accurately resolve between new physics models and the Standard
  Model. Furthermore, a lack of understanding of the precision of
  simplistic estimates may result in faulty conclusions.

Auto-generated calculations from \FS
\cite{Athron:2014yba,Athron:2017fvs,Athron:2021kve} and \sarah/\spheno
\cite{Staub:2009bi,Staub:2010jh,Staub:2012pb,Staub:2013tta,Porod:2011nf},
on the other hand, aim to provide high precision calculations
available for a very wide range of models.  Unfortunately until now
both of these codes implemented \MSbar/\DRbar\ calculations that
suffer from non-decoupling logarithms that can spoil the precision if the
the new physics particles are heavy.  We hope that the calculation presented
here and its availability within the \FS package will allow for more
stable and precise calculations to be performed without requiring
projects and papers focused on just developing dedicated calculations
in individual models. While a proper quantitative assessment of the
actual uncertainty is model dependent and we do not provide a thorough
analysis of the numerical uncertainty in applications, we still hope
that the existence of standard tools such as \sarah/\spheno and \FS
implementing calculations like the one presented here can reduce the
burden of this by generating the calculations for many different
models in a uniform way.

This paper is structured as follows: In \secref{sec:nondecoupling} we
discuss the non-decoupling problem present in previous $M_W$
calculations in \FS\ and other spectrum generators;
in \secref{sec:calculation} we
describe the improved calculation of $M_W$ in \FS.  In
\secref{sec:applications} we study the $M_W$ prediction different
\BSM\ models with \FS.  We conclude in \secref{sec:conclusions}.

\section{Non-decoupling problem in BSM calculations of the W boson
  pole mass}
\label{sec:nondecoupling}

Since version 2.0 the $W$ boson pole mass $M_W$ can be predicted in
\FS-generated \BSM\ spectrum generators \cite{Athron:2017fvs}, given
the experimental values 
for the $Z$ boson pole mass $M_Z$ and the Fermi constant $G_F$ as
input.%
\footnote{The calculation of $M_W$, given $G_F$ as input, is enabled
  by default in \FS, if the \BSM\ model contains all \SM\ particles
  and the \BSM\ gauge group contains the \SM\ gauge group as a factor.
  If these conditions are not fulfilled, $M_W$ is used as input.  To
  enforce the calculation of $M_W$, set \code{FSWeakMixingAngleInput =
    FSFermiConstant} in the corresponding \FS\ model file.  To enforce
  the use of $M_W$ as input, set \code{FSWeakMixingAngleInput =
    FSMassW}.  See Ref.\ \cite{Athron:2017fvs} for details.}
That calculation, like the ones in other public spectrum generators,
suffered from a non-decoupling problem which we describe
here. Readers interested in the current calculation in  \FS\ 2.7.0 may
skip to \secref{sec:calculation}, where the improved calculation is
presented in a self-contained way.

The non-decoupling problem is present in all \BSM\ calculations of $M_W$
which are based on a straightforward generalization of the \SM\
calculation in the \MSbar\ scheme of Ref.\ \cite{Degrassi:1990tu}, such as
in Ref.\ \cite{Pierce:1996zz} and in the codes \sarah/\spheno
\cite{Staub:2009bi,Staub:2010jh,Staub:2012pb,Staub:2013tta,Porod:2011nf}
and \FS\ 2.0.0--2.4.2~\cite{Athron:2017fvs}. The issue was not a problem as
long as the precision need for \BSM\ calculations was lower and as long
as only \BSM\ masses around the electroweak scale were considered. The
problem was first explicitly mentioned and discussed in
Ref.\ \cite{Diessner:2019ebm}, where the \sarah/\spheno \MRSSM implementation
was compared with an on-shell scheme calculation with manifest
decoupling.

In order to explain the problem and its origin we start from the basic
relation for $M_W$ put forward in Ref.\ \cite{Degrassi:1990tu},
updated in Ref.\ \cite{Degrassi:2014sxa},\footnote{%
  This equation is equivalent to Eqs.\ (57) and (67) of
  Ref.\ \cite{Athron:2017fvs} if the appropriate relationship between
  the quantities $\Delta\hat{r}$ and $\Delta\hat{r}_W$ is used.
  Here and in all of this section we ignore the possibility for tree-level contributions to the
  $\hat\rho$ parameter $\Delta\hat\rho_\tree\ne0$, since they are not relevant
  for the present discussion.}
\begin{align}
{G_F}
=
  \frac{\pi\,\aem(M_Z)}{\sqrt{2}\,M_W^2\,\left(1-\frac{M_W^2}{\hat\rho M_Z^2}\right)} 
\left(1+{{\Delta \hat{r}_W}}
\right)\,.
\label{DegrassiBasis}
\end{align}
The building blocks are defined in the original references; in
particular the \FS\ implementations are defined in
Refs.\ \cite{Athron:2014yba,Athron:2017fvs}. To illustrate the problem
we focus on the influence of the fine structure constant
\begin{align}
  \aem(M_Z)=\frac{\alpha_{\text{em,\SM}}^{(5),\MSbar}(M_Z)}{1-\Delta\aem^\SM-\Delta\aem^\BSM}\,,
\end{align}
where
$\Delta\aem^\BSM$ is a one-loop expression containing logarithms of the
form $\log(m_\BSM/M_Z)$.

In calculations using this approach, Eq.\ \eqref{DegrassiBasis} is solved
for $M_W$ and evaluated
exactly by numerical iteration, while its building blocks such as
$\Delta\aem^\BSM$ are evaluated at fixed order perturbation theory.
In the \SM\ it is known that this approach resums important higher-order
contributions \cite{Degrassi:1990tu}, similar to the resummation in
the context of the on-shell scheme of Ref.\ \cite{Consoli:1989fg}.

However, in a \BSM\ context with heavy \BSM\ masses the situation
changes. E.g.\ $\Delta\aem^\BSM$ schematically enters as
\begin{align}
{G_F}
=
\left[  \frac{\pi\,\aem(M_Z)}{\sqrt{2}\,M_W^2\,\left(1-\frac{M_W^2}{\hat\rho M_Z^2}\right)} 
\left(1+{{\Delta \hat{r}_W}}
\right)
\right]_{\Delta\aem^\BSM=0} \times \left[1+\Delta\aem^\BSM+(\Delta\aem^\BSM)^2+\cdots\right]\,,
\label{deltaaemterms}
\end{align}
where the dots denote further terms involving $\Delta\aem^\BSM$ of
two-loop and higher order.
For large \BSM\ masses, $\Delta\aem^\BSM$ increases logarithmically,
i.e.\ it is non-decoupling. It is clear that in a complete calculation
the prediction for $M_W$ such non-decoupling effects must cancel order
by order.

The calculations of Refs.\ \cite{Pierce:1996zz,Allanach:2001kg,Allanach:2013kza, Staub:2009bi,Staub:2010jh,Staub:2012pb,Staub:2013tta,Porod:2011nf,Athron:2017fvs} are complete at the one-loop \BSM\
level. Hence the term $\Delta\aem^\BSM$ in Eq.\ \eqref{deltaaemterms}
combines with other one-loop terms such that the large one-loop logarithms cancel
in the prediction for $M_W$. However the effectively generated
two-loop logarithms from the  $(\Delta\aem^\BSM)^2$-term
cannot cancel since the calculation is not complete at the two-loop level.

For this reason the numerical evaluation of Eq.\ \eqref{DegrassiBasis} with
one-loop evaluation of all \BSM\ building blocks leads to non-cancelling
large logarithms. These are formally of two-loop or higher order and arise via
$(\Delta\aem^\BSM)^2$ and similar terms involving the other building
blocks $\Delta\hat\rho$ and $\Delta \hat{r}_W$.

In typical applications such as the ones presented below or in
Ref.\ \cite{Diessner:2019ebm}, the numerical value of
$\Delta\aem^\BSM$ is around a few percent. Accordingly,
$(\Delta\aem^\BSM)^2$ and other non-decoupling two-loop effects can be
numerically estimated to be of the order permille and to shift $M_W$
by $\mathcal{O}(100\MeV)$. This is exactly what has been observed in
Ref.\ \cite{Diessner:2019ebm}, see in particular their \figurename~9. The
behavior of the \FS\ 2.0~computation of $M_W$
is essentially the same as the one of \sarah/\spheno, see \figurename~10 of
Ref.\ \cite{Athron:2017fvs}.

In order to avoid the non-decoupling problem, the computation must be
explicitly truncated such that \BSM\ effects are taken into account
precisely at one-loop order. The resulting approach is described in the
next section.

\section{Calculation of the W boson pole mass}
\label{sec:calculation}

In \FS\ 2.5.0 the calculation of $M_W$
has been modified to increase the precision and to avoid the non-decoupling
behavior explained in the previous section (which used to be
present e.g.\ in the calculation of $M_W$ performed by
\FS\ 2.0~\cite{Athron:2017fvs}).
The decoupling behavior is
achieved by a strict one-loop calculation of \BSM\ contributions, avoiding spurious
incomplete two-loop contributions, such that in the limit of
increasing \BSM\ particle masses the
predicted value for $M_W$ converges to the \SM\ prediction.  In \FS\ 2.7.0 the calculation of
$M_W$ has been refined further to include the state-of-the-art \SM\
contributions.

In the following we briefly describe the calculation
of $M_W$ in \FS\ 2.7.0.

\subsection[Calculation of $M_W$ in FlexibleSUSY]{Calculation of $\boldsymbol{M_W}$ in FlexibleSUSY}
\label{sec:calculation_MW_FS}

The calculation of $M_W$ in \FS\ since version 2.5.0 is an adaptation of the
procedure presented in Ref.~\cite{Degrassi:2014sxa} to \BSM\ models,
starting again effectively from Eq.\ \eqref{DegrassiBasis}. 
Its goals are to avoid the non-decoupling problem for heavy
\BSM\ masses but also to treat the \SM\ contributions with the highest
possible precision. Given the discussion of the previous section, the
solution is to use Eq.\ \eqref{DegrassiBasis}, split all its building
blocks into a sum of \SM\ and \BSM\ contributions, and solve for  the
ratio $M_W^2 /(M_W^\SM)^2 $, where $M_W^\SM$ is the
\SM\ prediction. This ratio is analytically evaluated as a 
strict fixed-order perturbative series, in our case truncated at
one-loop order. In this way, decoupling is obtained, and the
state-of-the-art \SM\ prediction can be combined with a fixed-order
\BSM\ correction.
As a result,
the $W$ boson pole mass can be expressed as
\begin{align}
  M_W^2 = (M_W^\SM)^2 \left(1 + \Delta_W\right).
  \label{eq:MW}
\end{align}
In \FS\ 2.5.0--2.6.2 the
value of $M_W^\SM$ has been fixed to $M_W^\SM = 80.385\GeV$, and since
version 2.7.0 the value of $M_W^\SM$ is calculated using the fit
formula of Eq.~(45) from Ref.~\cite{Degrassi:2014sxa}.
The term $\Delta_W$ contains the tree-level and one-loop \BSM\
contributions in the relation \eqref{eq:MW} between the \SM\
prediction $M_W^\SM$ and the \BSM\ prediction $M_W$ and is given by
\begin{align}
  \Delta_W = \frac{s_W^2}{c_W^2 - s_W^2} \left[
    \frac{c_W^2}{s_W^2}\left( \Delta\hat\rho_\tree + \Delta\hat\rho_\BSM \right)
    - \Delta\hat{r}_{W,\BSM} - \Delta\aem^\BSM
  \right].
  \label{eq:DeltaW}
\end{align}
All parameters entering this equation are consistently defined as
\BSM\ \MSbar/\DRbar\ parameters at the renormalization scale
$Q=M_Z$.\footnote{The renormalization scale used to calculate
  $\Delta_W$ with Eq.~\eqref{eq:DeltaW} is defined by the variable
  \lstinline|LowScale| in the corresponding \FS\ model file. By
  default the variable is set to \lstinline|LowScale = LowEnergyConstant[MZ]|,
  which corresponds to $Q=M_Z$.\label{foo:Q}}
Accordingly all appearing one-loop
corrections are evaluated with the
renormalization scale set to $Q=M_Z$.  We have abbreviated $s_W \equiv
\sin(\theta_W)$ and $c_W \equiv \cos(\theta_W)$, where $\theta_W$ is
the \BSM\ weak mixing angle in the \MSbar/\DRbar\ scheme at 
$Q=M_Z$. It is calculated from the relation \cite{Degrassi:1990tu}  
\begin{align}
  s_W^2 c_W^2 =
  \frac{\pi\,\aem(M_Z)}{\sqrt{2}\,M_Z^2\,G_F\,\hat\rho_\tree \left(1-\Delta\hat{r}\right)} .
  \label{eq:calculation_for_theta_W}
\end{align}
With $\aem(M_Z)$ we denote the \MSbar/\DRbar\ electromagnetic coupling
in the \BSM\ model at the renormalization scale $Q=M_Z$.  
The calculation of the loop correction $\Delta\hat{r}$ in
Eq.~\eqref{eq:calculation_for_theta_W} is described in Section~8 of
Ref.~\cite{Athron:2017fvs}.

The term $\Delta\hat\rho_\tree$ in Eq.~\eqref{eq:DeltaW} denotes the
tree-level \BSM\ contribution to the $\rho$-parameter and is defined
as
\begin{align}
  \Delta\hat\rho_\tree = \hat\rho_\tree - 1,
\end{align}
where $\hat\rho_\tree$ is the tree-level $\rho$-parameter in the \BSM\
model.\footnote{%
In this paper we always treat $\Delta\hat\rho_\tree$ as a small
correction which is at most of the same numerical order as
\BSM\ one-loop corrections. Accordingly, the notion of ``full
\BSM\ one-loop order'' includes $\Delta\hat\rho_\tree$, while products
of $\Delta\hat\rho_\tree$ and one-loop quantities are
neglected. Whenever we discuss decoupling and the limit of heavy
\BSM\ masses $m_{\text{\BSM}}$, we implicitly assume that potential \BSM\ vacuum
expectation values contributing to $\Delta\hat\rho_\tree$ behave as
$1/m_{\text{\BSM}}^2$.\label{foo:rho_hat_tree}
}
The term $\Delta\hat\rho_\BSM$ in Eq.~\eqref{eq:DeltaW} contains the
pure \BSM\ one-loop contributions in the relation between the \SM\
$\rho$-parameter, $\hat\rho_\SM$, and the loop-level \BSM\ $\rho$-parameter,
$\hat\rho_\BSM$, and is calculated as
\begin{align}
  \Delta\hat\rho_\BSM &=
  \frac{1}{m_Z^2} \left[ \Sigma_Z(m_Z^2) - \Sigma^\SM_Z(m_Z^2) \right]
  - \frac{1}{m_W^2} \left[ \Sigma_W(m_W^2) - \Sigma^\SM_W(m_W^2) \right],
  \label{eq:Delta_rho_hat_BSM}
\end{align}
where $\Sigma_Z(p^2)$ and $\Sigma_W(p^2)$ are the real parts of the
transverse components of the momentum-dependent
\MSbar/\DRbar-renormalized full one-loop \BSM\ $W$ and $Z$ boson
self-energies, evaluated at the squared momenta $p^2=m_W^2$ and
$p^2=m_Z^2$, respectively, where $m_W$ and $m_Z$ are the \BSM\
\MSbar/\DRbar $W$ and $Z$ boson masses.  The symbols
$\Sigma^\SM_Z(p^2)$ and $\Sigma^\SM_W(p^2)$ are the corresponding
\MSbar-renormalized \SM\ counterparts.  The subtraction of the \SM\
from the \BSM\ $W$ and $Z$ self-energies is performed numerically so
that mixing effects from new \BSM\ particles are correctly taken into
account.
The one-loop contribution $\Delta\hat{r}_{W,\BSM}$ in
Eq.~\eqref{eq:DeltaW} contains the pure \BSM\ one-loop contributions
to the relation \eqref{DegrassiBasis} between $M_W$ and $G_F$
and is calculated as
\begin{align}
  \Delta\hat{r}_{W,\BSM} &=
  \frac{1}{m_W^2}\left[
    \Sigma_W(0) - \Sigma^\SM_W(0) - \Sigma_W(m_W^2) + \Sigma^\SM_W(m_W^2)
  \right]
  + \delta_{\text{VB}}^\BSM ,
  \label{eq:Delta_rw_hat_BSM}
\end{align}
where $\delta_{\text{VB}}^\BSM$ contains the pure \BSM\ vertex and box
diagram contributions.
The term $\Delta\aem^\BSM$ in Eq.~\eqref{eq:DeltaW} denotes the
one-loop threshold correction for the electromagnetic coupling between
the \SM\ and the \BSM\ model, c.f.\ Section~5 in
Ref.~\cite{Athron:2017fvs}.
Since the r.h.s.\ of Eq.~\eqref{eq:MW} depends on the value of $M_W$,
the equation is solved iteratively using the experimentally measured
value for $M_W = 80.385\GeV$ as initial value.

We remark that the prediction of $M_W$ in Eq.~\eqref{eq:MW} is
complete at the one-loop level w.r.t.\ the \BSM\ contributions.  As
described above this includes potential mixing effects between
\BSM\ and \SM\ states. In this procedure mixing effects are accounted
when the one-loop \SM\ contributions in the fit formula for $M_W^\SM$
are cancelled through the numerical subtraction of the same
contributions in Eqs.~\eqref{eq:Delta_rho_hat_BSM} and
\eqref{eq:Delta_rw_hat_BSM}, such that the full one-loop pure
\BSM\ contributions remain.  At the same time higher order \SM\
contributions incorporated through the use of the \SM\ fit formula do
not include these mixing effects.

As stressed before, the calculation of $M_W$ in Eq.~\eqref{eq:MW} shows
a decoupling property where $\Delta_W$ in Eq.\ \eqref{eq:DeltaW}
behaves as $1/m_\BSM^2$ for increasing \BSM\ particle masses.  This is
achieved by the strict treatment of the \BSM\ loop corrections at
one-loop level, where spurious two-loop and higher order contributions
are avoided.  It is particularly important that the expression for
$\Delta_W$ is evaluated in terms of a consistent set of parameters,
perturbatively truncated at one-loop order. We have chosen the set of
fundamental \BSM\ \MSbar/\DRbar\ parameters at the renormalization
scale $Q=M_Z$. One subtlety is that in this scheme, the running gauge
couplings and $\theta_W$ differ from the corresponding \SM\ values,
and the differences can contain non-decoupling logarithms of
higher order. However, this does not spoil the decoupling
property of the full expression $\Delta_W$, which is mainly governed
by the $1/m_\BSM^2$ behavior, while any non-decoupling logarithms from
$\theta_W$ would only appear as a multiplicative factor to terms with
this behavior. It would be possible to employ different schemes for
the couplings entering Eq.~\eqref{eq:DeltaW}, but studies of such
alternatives are outside the scope of the present paper. We refer to
Ref.\ \cite{Kwasnitza:2020wli}, particularly Section~3.2, for a
similar discussion of the role of different parametrizations and the
danger of including fake logarithmically enhanced higher-order terms.

\subsection[Uncertainty of $M_W$]{Uncertainty of $\boldsymbol{M_W}$}
\label{sec:calculation_uncertainties}

The uncertainty $\Delta M_W$ of the prediction of $M_W$ with
Eq.~\eqref{eq:MW} can be divided into two contributions: The
uncertainty $\Delta M_W^\SM$ from the calculation of $M_W^\SM$ from
Ref.~\cite{Degrassi:2014sxa}, which should be constructed from the
parametric uncertainty and the missing higher order corrections,
estimated at $9\MeV$ and $3\MeV$ respectively in
Ref.\ \cite{Degrassi:2014sxa}. Since we implement the \SM\ prediction
using the \MSbar\ fit formula there is also in principle an uncertainty from
the fit itself, but this should always be less than $0.5\MeV$.  A
second contribution is the uncertainty $\Delta M_W^\BSM$ from missing
higher order \BSM\ contributions.  The latter of the two uncertainties is
model-dependent and can be estimated for example by renormalization
scale variation.

There exist simple parametrizations of $M_W^\SM$ from
Eq.~\eqref{eq:MW} from \OS\ and \MSbar\ calculations.  These
parametrizations depend on the deviation of the masses $M_h$, $M_t$,
$M_Z$, the hadronic contributions to the fine structure constant
$\Delta\alpha^{(5)}_{\text{had}}$, and the strong coupling constant
$\as$ from predefined values.  The parametrization of $M_W^\SM$ from
the \OS\ calculation is given in Eqs.~(6)--(9) in Ref.\
\cite{Awramik:2003rn}.  The parametrization from the \MSbar\
calculation is given by Eq.~(45) in Ref.\ \cite{Degrassi:2014sxa}.
Both are implemented in \FS\ 2.7.0, which takes \FS's prediction of
the Higgs boson pole mass and uses it to calculate the $W$ boson pole
mass.  We show the variation that occurs in both \OS\ and \MSbar\
predictions for $M_W^\SM$ in \tabref{tab:UncertaintyMW}.  As can been
seen in the table we get a variation of up to $\Delta M_W^\SM = 6\MeV$
when the top quark pole mass is varied over a $1\GeV$ range (c.f.\ the
discussion of the uncertainty and the ambiguity of the definition of
$M_t$ in Section~60 of Ref.~\cite{ParticleDataGroup:2020ssz}).

\begin{table}[tb]
  \begin{center}
    \begin{tabular}{lrr}
      \toprule
      & $\Delta M_W^\SM$ (\OS) & $\Delta M_W^\SM$ (\MSbar) \\
      \midrule
      $M_h \pm 1\sigma$ & $0.08\MeV$ & $0.08\MeV$ \\
      $M_t \pm 1\sigma$ & $1.8\MeV$  & $1.8\MeV$ \\
      $M_t \pm 1\GeV$   & $6.0\MeV$  & $6.1\MeV$ \\
      $M_Z \pm 1\sigma$ & $2.6\MeV$  & --- \\
      $\Delta\alpha^{(5)}_{\text{had}} \pm 1\sigma$ & $1.3\MeV$ & $1.3\MeV$ \\
      $\as \pm 1\sigma$ & $0.59\MeV$ & $0.62\MeV$ \\
      \bottomrule
    \end{tabular}
    \caption{Variation of the $M_W^\SM$ pole mass, calculated in the
      on-shell and \MSbar\ scheme, respectively.  $M_h$, $M_t$, $M_Z$,
      $\Delta\alpha^{(5)}_{\text{had}}$, $\as$ are varied around their
      central values given in Eqs.~\eqref{eq:SM_parameters} over their
      $1\sigma$ range from the 2021 update of
      Ref.~\cite{ParticleDataGroup:2020ssz}, and $M_t$ is also varied
      by $1\GeV$.  Note that since the \MSbar\ fit formula from
      Ref.~\cite{Degrassi:2014sxa} is independent of $M_Z$ we do not
      give a corresponding variation of $M_W^\SM$.}
    \label{tab:UncertaintyMW}
  \end{center}
\end{table}

Thanks to the decoupling property of the calculation, the prediction
of  $M_W$ is precise even for heavy \BSM\ spectra,
where the uncertainty of the $M_W$ prediction is only dominated by the
\SM\ prediction, i.e.\ $\Delta M_W \approx \Delta M_W^\SM$.  We study
and illustrate this decoupling property in \secref{sec:applications}
for concrete \BSM\ models.

\section{Applications}
\label{sec:applications}

In the following we study the $W$ boson pole mass prediction in
different \BSM\ models with \FS, as described in
\secref{sec:calculation}.  We focus in particular on potential
deviations from the \SM\ prediction.  If not stated otherwise, we use
the following values for the \SM\ parameters from
Ref.~\cite{ParticleDataGroup:2020ssz}:
\begin{align}
  \aem(M_Z) &= 1/127.916, &
  G_F &= 1.1663787\times 10^{-5}\GeV^{-2}, &
  \as(M_Z) &= 0.1179, \nonumber \\
  M_Z &= 91.1876\GeV, &
  M_t &= 172.76\GeV, &
  m_b(m_b) &= 4.18\GeV,
  \label{eq:SM_parameters} \\
  \Delta\alpha^{(5)}_{\text{had}} &= 0.02766. \nonumber
\end{align}
When the Higgs mass is also fixed to its measured value,
$M_h=125.25\GeV$, in addition to using the parameters in
Eqs.~\eqref{eq:SM_parameters} for the other inputs, the \SM\
prediction for the $W$ boson pole mass in the \MSbar\ scheme is
$M_W^\SM = 80.351\GeV$ \cite{Degrassi:2014sxa}.
This is the \SM\ prediction we use in the
following sections.\footnote{The on-shell calculation from
  Ref.~\cite{Awramik:2003rn} yields $M_W^\SM=80.355\GeV$ for the
  parameter set given in Eqs.~\eqref{eq:SM_parameters}, which is
  $4\MeV$ larger than the \MSbar\ calculation.  A more recent on-shell
  calculation with \FeynHiggs\ gave $M_W^\SM=(80.353 \pm 0.004)\GeV$
  \cite{Bagnaschi:2022qhb}.}


\subsection{MSSM}
\label{sec:MSSM}
The \MSSM\ is one of the most widely studied and best motivated
extensions of the Standard Model.  Therefore we demonstrate our new
calculation in this model. We use the following parameter scenario
with large stop and sbottom mass splitting and a common supersymmetry
mass scale $\MSUSY$:
\begin{align}
  \tan\beta &= 20, & \mu &= m_A = M_i = \MSUSY~(i=1,2,3), \nonumber \\
  A_t &= X_t + \mu/\tan\beta, & A_f &= 0~(f=u,d,c,s,b,e,\mu,\tau), \label{eq:MSSM_scenario} \\
  m^2_{\tilde{q}} &= m^2_{\tilde{u}} = m^2_{\tilde{d}} = \MSUSY^2\unity, &
  m^2_{\tilde{l}} &= m^2_{\tilde{e}} = (\MSUSY/2)^2\unity, \nonumber
\end{align}
with $X_t = -\sqrt{6}\MSUSY$.  The slepton masses are chosen to be
smaller than the squark masses in order to increase the loop
corrections to $M_W$.  The \SM-like Higgs boson pole mass $M_h$ has
been calculated using a fixed-order \DRbar\ calculation at the full
one-loop level, including dominant two-loop
\cite{Degrassi:2001yf,Brignole:2001jy,Dedes:2002dy,Brignole:2002bz,Dedes:2003km}
and three-loop contributions via the \Himalaya\ package
\cite{Harlander:2008ju,Kant:2010tf,Kunz:2014gya,Harlander:2017kuc}.

\begin{figure}[tb]
 \centering
 \includegraphics[width=0.49\textwidth]{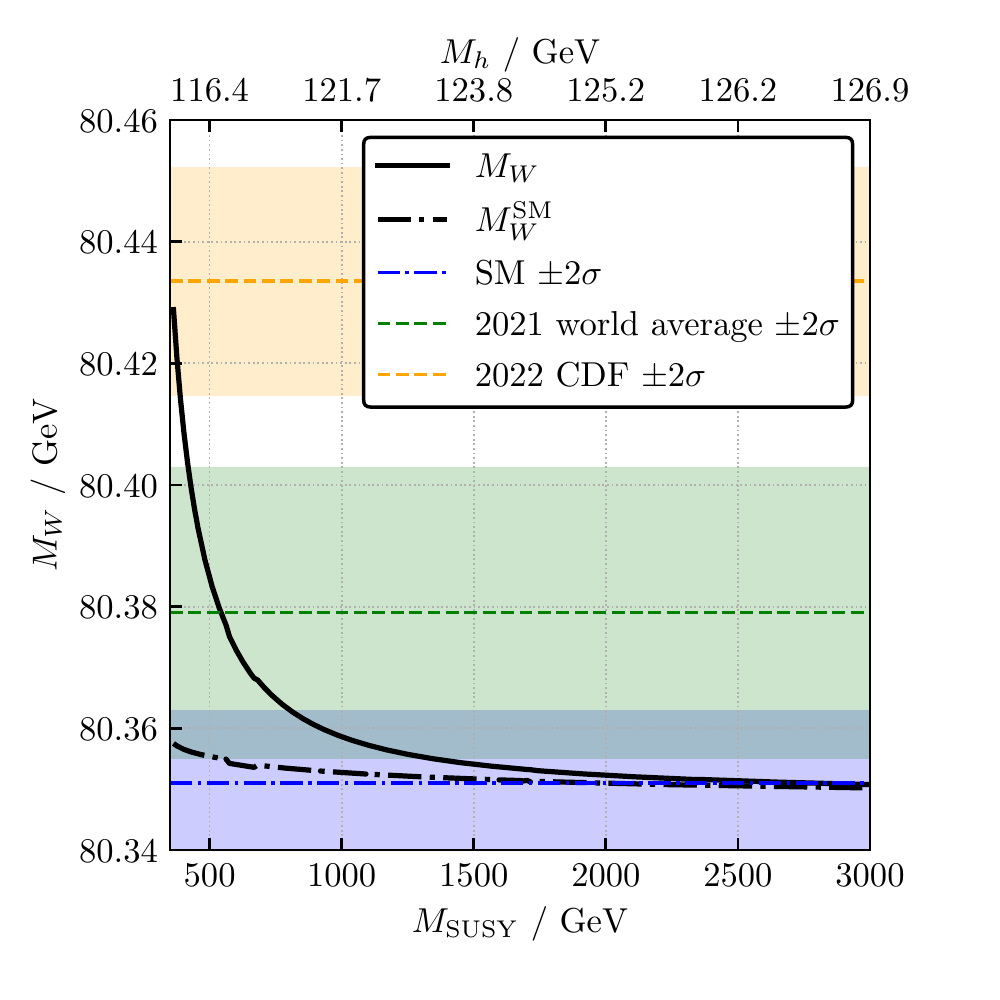}
 \caption{Prediction of $M_W$ in the \MSSM\ as a function of the
   common supersymmetry scale $\MSUSY$ (black solid line).  The black
   dash-dotted line shows the \SM\ prediction using the \MSbar\ fit
   formula from Ref.~\cite{Degrassi:2014sxa} for the corresponding
   value of the Higgs pole mass $M_h$ (top axis).  The blue
   dash-dotted line shows the \SM\ prediction
   $M_W^\SM=(80.351\pm 0.006)\GeV$ for the fixed value of the Higgs
   pole mass $M_h=125.25\GeV$, including its uncertainty.  The green
   dashed line shows the 2021 world average experimental value
   $M_W^{2021}=(80.379 \pm 0.012)\GeV$ including its uncertainty. The
   orange dashed line shows the \CDF\ value
   $M_W^\CDF=(80.4335 \pm 0.0094)\GeV$ including its uncertainty.}
 \label{fig:MW_MSSM}
\end{figure}

The value of $M_W$ predicted in the scenario \eqref{eq:MSSM_scenario} is shown in
\figref{fig:MW_MSSM} as black solid line as a function of the common
supersymmetry mass scale $\MSUSY$.  For increasing $\MSUSY$ the
prediction for $M_W$ converges to the \SM\ value $M_W^\SM$ (black
dash-dotted line), which nicely illustrates the decoupling behavior
of the calculation.
In these scenarios the experimental value for $M_h=125.25\GeV$ is
correctly predicted by the \MSSM\ for $\MSUSY \approx 2023\GeV$.  For
this value of \MSUSY\ the \MSSM\ predicts $M_W\approx 80.352\GeV$,
which deviates by $1\MeV$ from the \SM\ prediction of $M_W^\SM\approx
80.351\GeV$.  However in the simple scenario shown here we only see
significant \BSM\ corrections for light SUSY masses that are
excluded by experimental searches and where the \SM-like Higgs mass is
also too small to match the measured value.  Therefore the \CDF\ value $M_W^\CDF$
(orange band in \figref{fig:MW_MSSM}) cannot be explained in this
simple \MSSM\ scenario we have used to demonstrate the
decoupling property of our new calculation.

It is certainly possible to move away from this simplified scenario
with common SUSY masses to try to obtain larger values of $M_W$ when
the Higgs boson pole mass is $M_h=125.25\GeV$.  However we find it is rather
challenging to fit a very large value while consistently evading
experimental constraints on the sparticles and fitting the Higgs mass
measurement.  This is consistent with the findings in the literature
\cite{Heinemeyer:2013dia}, and a very recent paper
\cite{Bagnaschi:2022qhb} also showed that if one requires, in
addition, an explanation of muon $g-2$ data,  the maximum $M_W$ in
the \MSSM\ is $80.376\GeV$.  Therefore while we do not entirely exclude
the possibility of an \MSSM\ explanation we leave that for dedicated
studies.\footnote{In fact while this paper was in preparation a Ref.\ \cite{Yang:2022gvz} presented MSSM scenarios with large $M_W$ corrections fitting the 2022 \CDF\ measurement.} However we expect such scenarios will be rare and conclude that the
prospects for explaining the 2022 \CDF\ measurement are not great, and if
it is confirmed this measurement motivates other extensions where
larger $M_W$ corrections are easier to obtain.


\subsection{Scalar Singlet Model}
\label{sec:SSM}

Before we consider a model which can give larger corrections to the $W$ boson pole mass we 
first briefly demonstrate the calculation in one of the most popular non-SUSY 
models, the Scalar Singlet Model (\SSM).  The \SSM\ is a simple 
extension of the \SM\ which couples a real scalar to the Higgs doublet. Here we 
look at the $\mathbb{Z}_2$ symmetric \SSM, which restricts the allowed 
interactions between the Higgs and scalar singlet, giving an extended Higgs potential of
\begin{equation} \label{eq:SSM_H_Potential}
 V  = -\mu^2 |H|^2 + \frac{\lambda}{2} |H|^4 + \frac{\mu_S^2}{2} S^2 + \frac{\lambda_S}{2} S^4 + \frac{\lambda_{HS}}{2} H^\dagger H S^2.
\end{equation}
The coupling $\lambda_{HS}$ allows for mixing between the neutral
component of the Higgs doublet $H$ and the neutral scalar singlet when
both have vacuum expectation values (VEVs) $v$ and $v_S$, respectively,
\begin{align}
  H &= \begin{pmatrix}\sigma^\pm \\ \frac{1}{\sqrt{2}} \left(\phi^0 + v + \imath \sigma^0\right)\end{pmatrix} &
  &\text{and} & 
  S &= s + v_S,
\end{align}
which we assume here.  Thus, after electroweak symmetry breaking
(EWSB), the neutral scalar fields $\phi_0$ and $s$ are rotated into
the Higgs boson mass eigenstates $(h_1, h_2)^T$ via the mixing matrix
$R(\alpha)$,
\begin{align} 
  \begin{pmatrix} h_1 \\ h_2 \end{pmatrix} &= R(\alpha) \begin{pmatrix} \phi^0 \\ s \end{pmatrix}, &
  &\text{where} &
  R(\alpha) &= \begin{pmatrix} \cos(\alpha) & -\sin(\alpha) \\ \sin(\alpha) & \cos(\alpha) \end{pmatrix}.
\end{align}
The \SM-like Higgs mass is determined via the mixing angle $\alpha$.  If 
$\sin^2(\alpha) < 0.5$ the lighter Higgs is doublet-dominated and thus is 
associated with the \SM-like Higgs.  If instead $\sin^2(\alpha) > 0.5$, then 
the lighter Higgs is singlet-dominated.\footnote{Note that this selection is 
	not done automatically in \FS~2.7.0.  Instead the user must specify which index of
	the Higgs mass eigenstate multiplet corresponds to the \SM-like Higgs.  For 
	example, this can be set in entry 22 of the \lstinline|FlexibleSUSY| block 
	in a Les Houches input file.  In this model we fixed entry \lstinline|FlexibleSUSY[22]| 
	to $0$ if the point has \lstinline|ZH[1,2]| $<\sqrt{0.5}$ (i.e.\ the entry of 
	the block \lstinline|ZH| corresponding to an off-diagonal element of 
	$R(\alpha)$ or $\sin^2(\alpha)$ is less than $0.5$), 	and 
	\lstinline|FlexibleSUSY[22]| fixed to $1$ if \lstinline|ZH[1,2]| $>\sqrt{0.5}$.}
In this model when the singlet-dominated state is heavier than the \SM-like Higgs, we 
expect negative corrections to the $W$ boson pole mass \cite{Lopez-Val:2014jva}.

\begin{figure}[tb]
	\centering
	\includegraphics[width=0.49\textwidth]{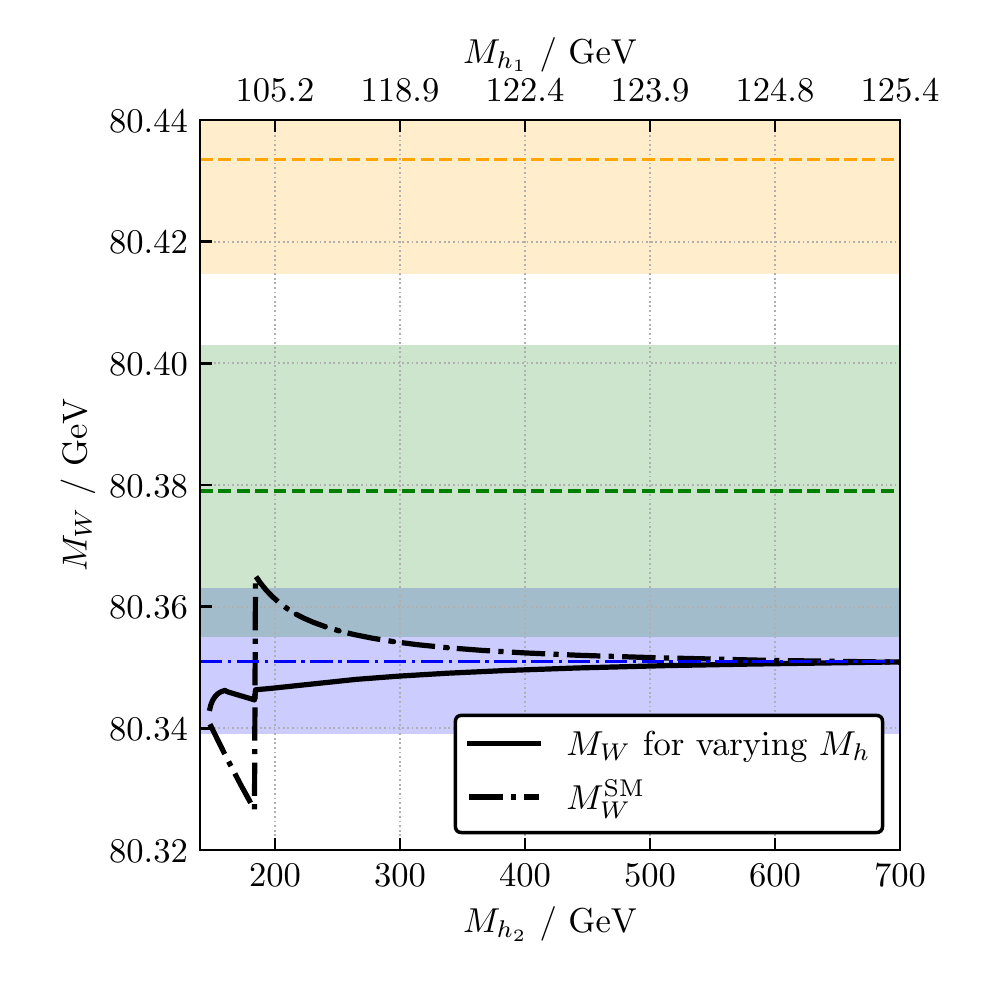}\hfill
	\includegraphics[width=0.49\textwidth]{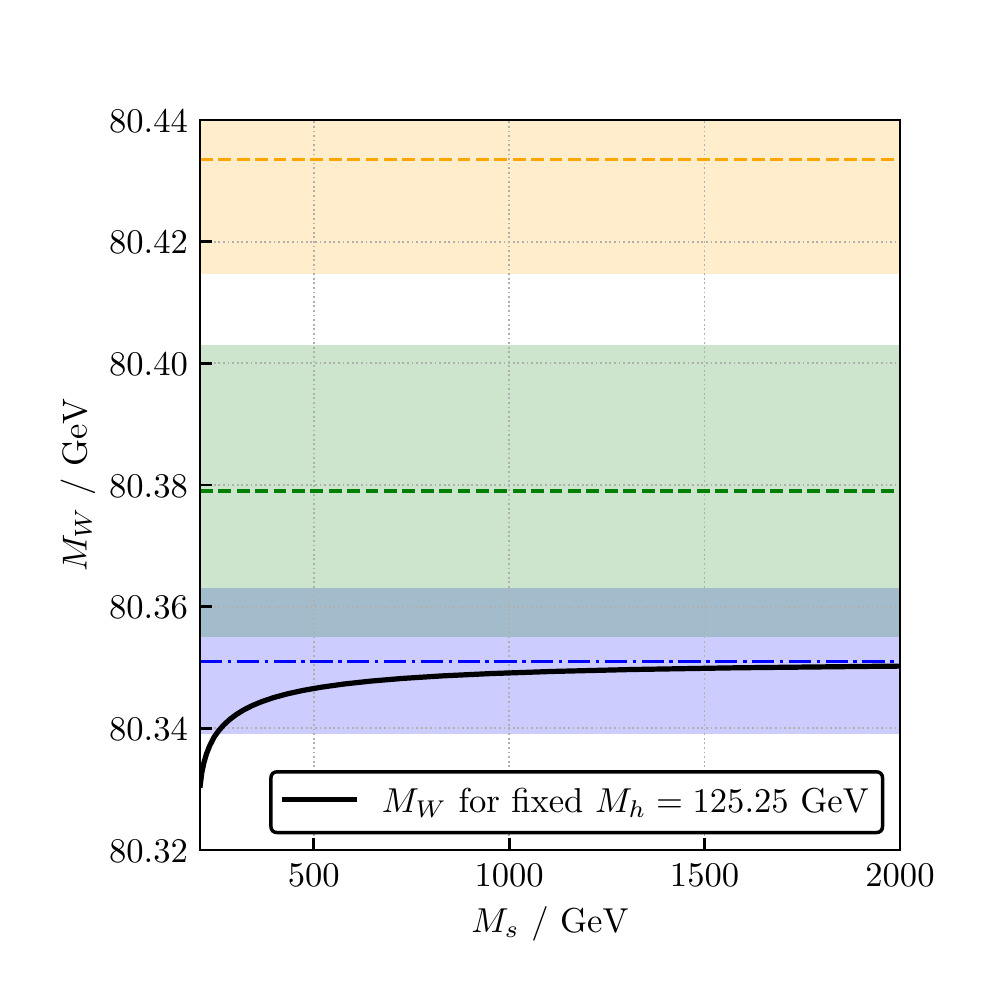}
	\caption{Prediction of $M_W$ in the the Scalar Singlet Model with a
		$\mathbb{Z}_2$ symmetry.  The blue, green and orange lines and regions are
		as in \figref{fig:MW_MSSM}.  In the left panel, the parameter set
		\eqref{eq:SSMparameters_a} is used.  As in \figref{fig:MW_MSSM},
		the solid black lines indicate the prediction of $M_W$ where the
		Higgs mass has been allowed to vary, and the black dash-dotted
		line shows the \SM\ prediction using Eq.~(45) from
		Ref.~\cite{Degrassi:2014sxa}.  The light and heavier Higgs masses
		are shown on the top and bottom axes, respectively.  The right panel
		uses the parameter set from \eqref{eq:SSMparameters_b} and shows
		the prediction of $M_W$ for a Higgs boson pole mass fixed to
		$M_h=125.25\GeV$, as a function of the scalar mass $M_s$, which is
		shown on the bottom axis.}
	\label{fig:MW_SSM}
\end{figure}

Following the Lagrangian of the \SSM\ we examine two sets of parameter choices,
\begin{subequations}
\begin{align}
	\lambda &= 0.36, & \lambda_S &= 1.4, & \lambda_{HS} &= 0.8, \label{eq:SSMparameters_a} \\
	M_h &= 125.25\GeV, & \lambda_S &= 2.122, & \lambda_{HS} &= 0.9917, \label{eq:SSMparameters_b}
\end{align} \label{eq:SSMparameters}%
\end{subequations}
while $\mu_S^2$ and $\mu^2$ are fixed to fulfill the one-loop
electroweak symmetry breaking conditions.  We scan over $v_S$, and get
the results shown in \figref{fig:MW_SSM}.  The left panel is included
to show how mixing effects between the SM Higgs and the scalar singlet
are handled and subtle issues related to this.  Here the input parameters
are fixed as in Eqs.\ \eqref{eq:SSMparameters_a}, which
includes the Higgs quadratic coupling $\lambda$ and leaves both Higgs
masses $M_{h_i}$ varying (rather than fixing one to $125.25\GeV$).
The left panel shows two distinct regions, separated by a
discontinuity in the \SM\ contribution to $M_W$ shown as the
dash-dotted black line. The boundary of the two regions is at
approximately $v_S\approx 64\GeV$ and $M_{h_2}\approx 184\GeV$.
On the
left side of this boundary the singlet-dominated pole mass $M_s$ is
smaller than the \SM-like Higgs mass $M_h$, while on the right side it
is larger.  Note that the \SM\ contribution to $M_W^\SM$ is varying because it
depends on the \SM-like Higgs mass, which is not fixed to the measured
value in this plot, but is instead varying. The mixing angle
$\alpha$ is used to determine which state is the \SM-like one.  As
$\sin(\alpha)$ passes through $\sqrt{0.5}$, the mass of the state we
treat as \SM-like changes from $M_h = M_{h_1}\approx 98\GeV$ to
$M_h = M_{h_2}\approx184\GeV$, leading to the discontinuity in
the \SM\ contribution $M_W^{\SM}$.

However, as described in \secref{sec:calculation}, this mixing effect
is correctly accounted for at the one-loop level in our calculation of
$M_W$ in Eq.~\eqref{eq:MW}.  This is achieved by the inclusion of the
full \BSM\ one-loop contributions and the numerical subtraction of the
pure \SM\ one-loop part in $\Delta_W$.  In this way the discontinuity
is avoided in our \SSM\ calculation of $M_W$ at the one-loop level
(black solid line in the left panel).  However, the \SM\ Higgs mass
also enters the calculation of $M_W^{\SM}$ at the two-loop level,
which is incorporated in the \MSbar\ fit formula we use.  This
two-loop pure \SM-Higgs part is not cancelled in our calculation,
because we aim to correctly include all known higher order corrections
to the \SM\ contribution (via the fit formula) and do not include
\BSM\ contributions beyond one-loop.

Nonetheless, in the case of mixing between \BSM\ and \SM\ states, such
as the Higgs as we see here, there is no clear definition of which
state is the \SM\ one.  Therefore when we fix the \SM\ Higgs to be the
state which has the most doublet content, as we do here, we find a
small discontinuity in the prediction of $M_W$ in the \BSM\ model,
which is of two-loop order.  This effect can be seen in the \SSM\
studied here in the small discontinuity in the solid black line in the
left panel in \figref{fig:MW_SSM}. The impact of this is
of the order $1$--$2\MeV$ and is thus encouragingly
small, suggesting that the impact from this mixing issue is not large.

When the singlet-dominated state is the lighter one, we find positive
contributions to $M_W$ compared to the \SM\ contributions, though for
this unrealistic choice of the Higgs mass the prediction is actually
smaller than that of the \SM\ with $M_h = 125.25\GeV$, as indicated by
the blue dash-dotted line in left panel of \figref{fig:MW_SSM}.
When $v_S>64\GeV$, the singlet state is
heavier than the \SM-like Higgs state so the contributions are negative.
As $v_S$ increases towards the right of the plot the scalar singlet
effectively decouples from the \SM, and we see that the \SSM\ and
\SM\ values of $M_W$ approach each other.  In the right panel of \figref{fig:MW_SSM}, where
the Higgs mass has been fixed to $M_h = 125.25\GeV$, we examine the
case $M_s > M_h$.  This shows that while in the \SSM\ one can get
non-negligible negative contributions when the two Higgses are close
together in mass, as the mass scale of the scalar increases it
eventually decouples leaving behind the \SM\ value of $M_W$.

We do not expect that the \SSM can explain the value of $M_W$ from the
2022 \CDF\ measurement, because large positive corrections require a
singlet-dominated state that is lighter than the \SM-like Higgs, with very
large singlet-doublet mixing.  However it is clear that $M_W$ is a
relevant constraint in this model and the precision calculation
presented here can be used in precision tests of the Scalar Singlet Model.


\subsection{MRSSM}
\label{sec:MRSSM}

The \MRSSM is a non-minimal supersymmetric extension of the Standard
Model which has \BSM\ contributions to the $W$ boson mass at
tree-level that can increase it above the \SM\ prediction.  Due to its
$R$-symmetry all gauginos are Dirac particles in the \MRSSM, and
unwanted sources of flavor violation from supersymmetry-breaking
trilinear couplings are forbidden.  A detailed description of the
model and the precise definitions for all parameters we refer to here
can be found in Section~2 of Ref.~\cite{Diessner:2014ksa}.

In addition we highlight here that the \MRSSM\ contains two specific
mechanisms which can increase the $W$ boson mass, which distinguish it
from e.g.\ the MSSM. The first mechanism is the appearance of new
Yukawa-like parameters in the superpotential $\Lambda_{u,d}$,
$\lambda_{u,d}$ which contribute to the $W$ boson pole mass similarly
as the top-Yukawa coupling \cite{Diessner:2014ksa}.  Secondly, the
\MRSSM\ contains a Hypercharge $Y=0$, $SU(2)_L$ Higgs triplet $T$ with
vacuum expectation value $v_T$, which is responsible for the positive
tree-level contribution to the $W$ boson mass
\cite{Kribs:2007ac,Diessner:2014ksa,Diessner:2019ebm}.

At tree-level the relation between the \DRbar\ $W$
and $Z$ boson masses $m_W$ and $m_Z$ reads,
\begin{align}
  m_W^2 = m_Z^2 \cos^2 \theta_W + g_2^2 v_T^2,
\end{align}
where $\tan \theta_W \equiv g_1/g_2$, which corresponds to a tree-level $\rho$-parameter of
\begin{align}
  \hat\rho_\tree = 1 + \frac{4v_T^2}{v_d^2 + v_u^2}
\end{align}
with $g_1$ and $g_2$ being the $U(1)_Y$ and $SU(2)_L$ \DRbar\ gauge
couplings, respectively.
\begin{figure}[tb]
 \centering
 \includegraphics[width=0.49\textwidth]{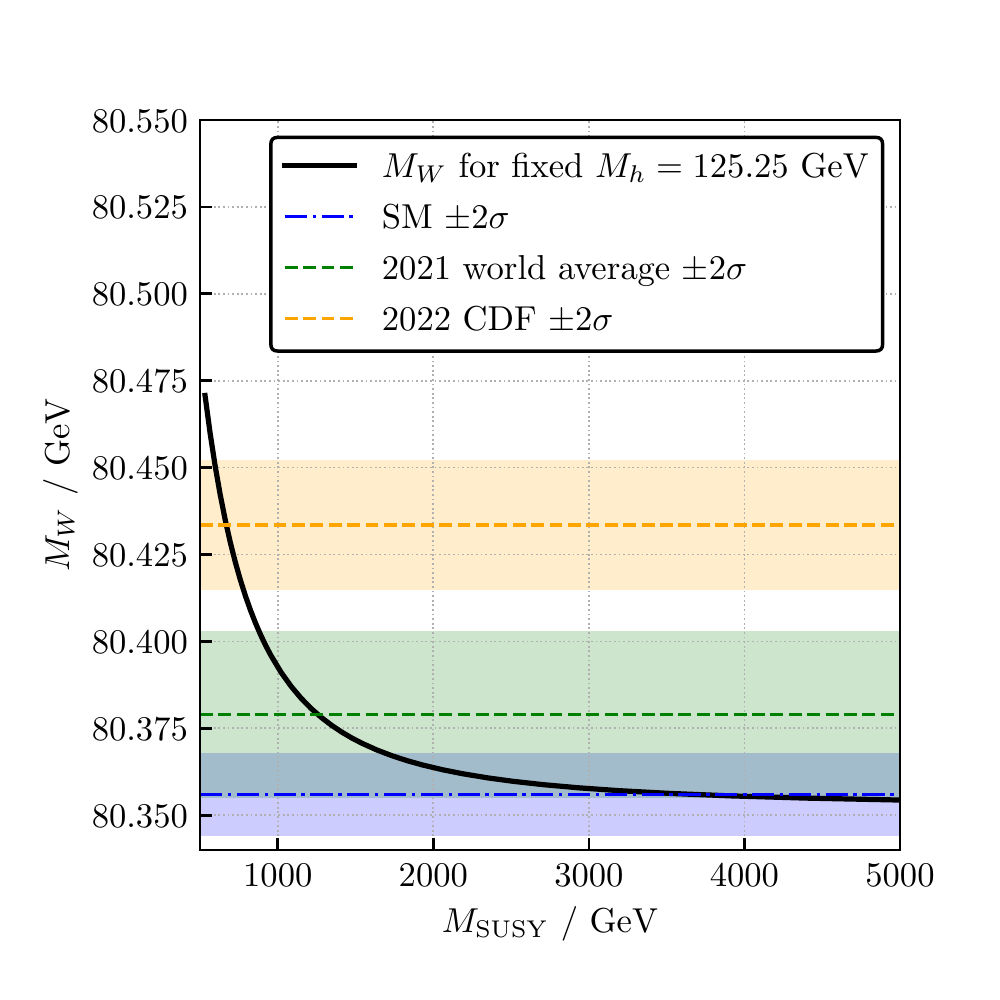}\hfill
 \includegraphics[width=0.49\textwidth]{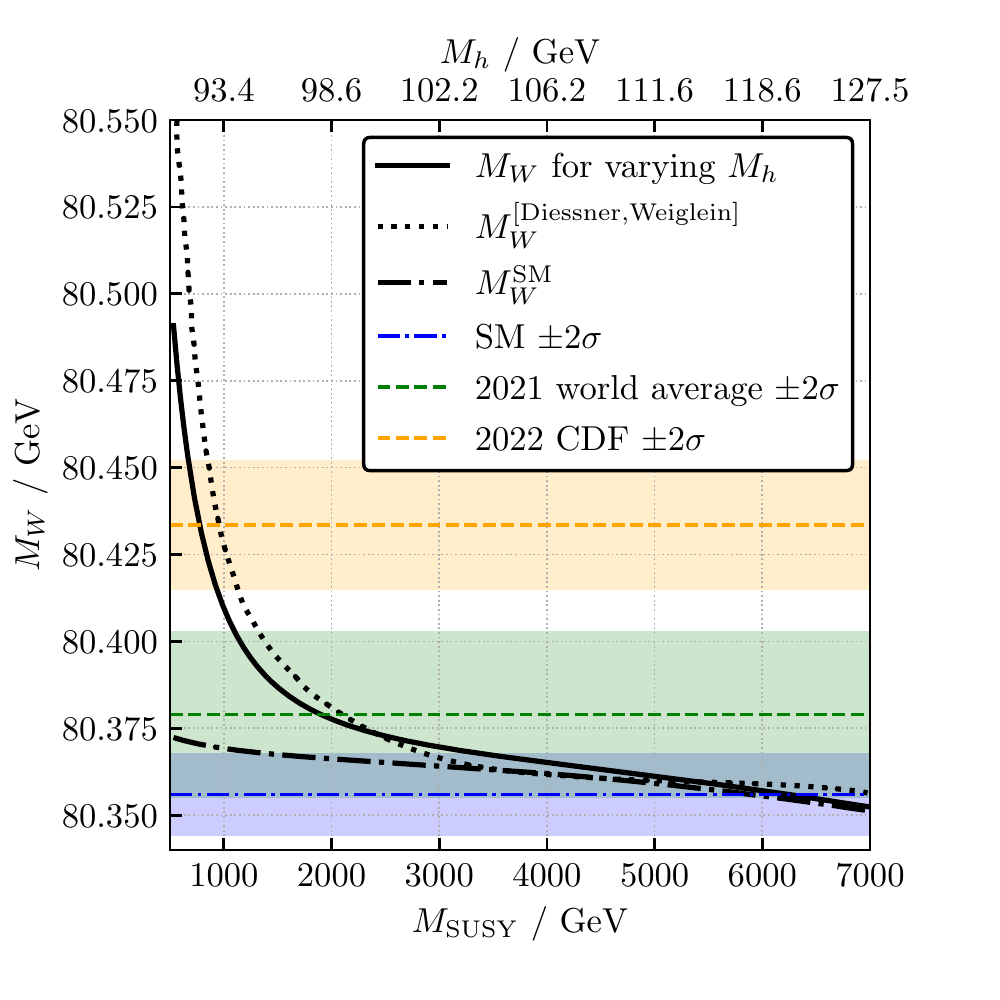}
 \caption{Prediction of $M_W$ in the \MRSSM\ as a function of the
   common supersymmetry scale $\MSUSY$ (black solid line), for
   parameter values as given in Eq.\ \eqref{eq:MRSSM_scenario} and below.  In the
   left panel the Higgs pole mass that is used to calculate $M_W^\SM$
   is fixed to $M_h=125.25\GeV$.  In the right panel the Higgs pole
   mass as predicted by the \MRSSM\ is used to calculate $M_W^\SM$.
   The black dash-dotted line shows the \SM\ prediction using the
   \MSbar\ fit formula from Ref.~\cite{Degrassi:2014sxa} for the
   corresponding value of the Higgs pole mass $M_h$ (top axis).  The
   black dotted line shows the prediction from
   Ref.~\cite{Diessner:2019ebm}.  The blue dash-dotted line shows
   the \SM\ prediction $M_W^\SM=(80.354\pm 0.006)\GeV$ for the fixed
   value of the Higgs pole mass $M_h=125.25\GeV$, including its
   uncertainty.  The green dashed line shows the 2021 world average
   experimental value $M_W^{2021}=(80.379 \pm 0.012)\GeV$ including
   its uncertainty. The orange dashed line shows the \CDF\ value
   $M_W^\CDF=(80.4335 \pm 0.0094)\GeV$ including its uncertainty.}
 \label{fig:MW_MRSSM}
\end{figure}
In \figref{fig:MW_MRSSM} we show the $W$ boson pole mass as predicted
by our improved calculation in the \MRSSM\ as a function of a common
supersymmetry mass scale \MSUSY, where the dimensionful superpotential
and soft-breaking \MRSSM\ parameters are set to
\cite{Diessner:2019ebm}
\begin{align}
  \begin{split}
    m^2_{R_u}=m^2_{R_d}=m_S^2=m_T^2=m_O^2=\frac{2 B_\mu}{\sin 2\beta} &= \MSUSY^2,\\
    m^2_{\tilde q}=m^2_{\tilde l} =m^2_{\tilde e}=m^2_{\tilde u}=m^2_{\tilde d} &= \MSUSY^2\unity, \\
    M^D_B=M^D_W=M^D_O=\mu_d=\mu_u &= \frac{\MSUSY}{2}.
  \end{split}
  \label{eq:MRSSM_scenario}
\end{align}
The dimensionless superpotential couplings are set to $\Lambda_d=-1$,
$\Lambda_u=-1.03$, $\lambda_d=1.0$ and $\lambda_u=-0.8$.  The ratio of
the Higgs doublet vacuum expectation values has been set to
$\tan\beta\equiv v_u/v_d=3$.  The Higgs triplet vacuum expectation
value $v_T$ is fixed by the one-loop electroweak symmetry breaking
conditions. The tree-level expressions for these can be found in
Eqs.~(2.15)--(2.16) in Ref.~\cite{Diessner:2014ksa}.  The top quark
pole mass and the strong coupling have been set to $M_t=173.0\GeV$ and
$\as(M_Z)=0.1181$, respectively, for comparison with
Ref.~\cite{Diessner:2019ebm}.

In the left panel of \figref{fig:MW_MRSSM} the \SM-like Higgs pole
mass $M_h$ that is used to calculate $M_W^\SM$ in Eq.~\eqref{eq:MW} is
fixed to the constant value $M_h=125.25\GeV$.  The value of $M_W$
predicted by the \MRSSM\ in this scenario (black solid line) shows the
described decoupling behavior for increasing \MSUSY\ and converges to
the \SM\ prediction with a constant offset due to $\hat\rho^\tree\neq
1$.  However, the \SM-like Higgs pole mass predicted by the
\MRSSM\ (using the one-loop fixed order calculation of \FS)
actually depends on \MSUSY\ and thus cannot be fixed to a constant
value.  To illustrate the effect of the variation of $M_h$ with
\MSUSY, we show in the right panel the prediction of $M_W$, where the
value of $M_h$ predicted by the \MRSSM\ is used to calculate
$M_W^\SM$.  Since $M_h$ is increasing with increasing \MSUSY (see
second horizontal axis on the top of the right panel), the value for
$M_W^\SM$ is no longer constant and decreases for increasing
\MSUSY\ (black dash-dotted line).  Due to the decoupling behavior,
the value of $M_W$ still converges to $M_W^\SM$.

In the shown scenario \eqref{eq:MRSSM_scenario} the experimental value
for $M_h=125.25\GeV$ is correctly predicted by the \MRSSM\ for $\MSUSY
\approx 6767\GeV$.  For this value of \MSUSY\ the \MRSSM\ predicts
$M_W\approx 80.353\GeV$, which deviates by $1\MeV$ from the
\SM\ prediction of $M_W^\SM\approx 80.352\GeV$, nicely illustrating
the decoupling behavior. However, in the shown scenario the value of
$M_W^\CDF$ (orange band in \figref{fig:MW_MRSSM}) cannot be reproduced
for realistic scenarios where the \SM\-like Higgs pole mass is close
to $125.25\GeV$.\footnote{It is well known that the fixed-order
  calculation of the Higgs mass we use here has a substantial
  uncertainty at large $\MSUSY$, which is around $\Delta M_h\approx 8\GeV$ when the
  SUSY scale is $7\TeV$ \cite{Athron:2016fuq}.  However even taking this
  sizeable uncertainty into account it is clear that in this particular
  scenario it is impossible to explain both measurements
  simultaneously.\label{foo:MhMW}}

\begin{figure}[tb]
 \centering
 \begin{subfigure}{0.49\linewidth}
 \includegraphics[width=\textwidth]{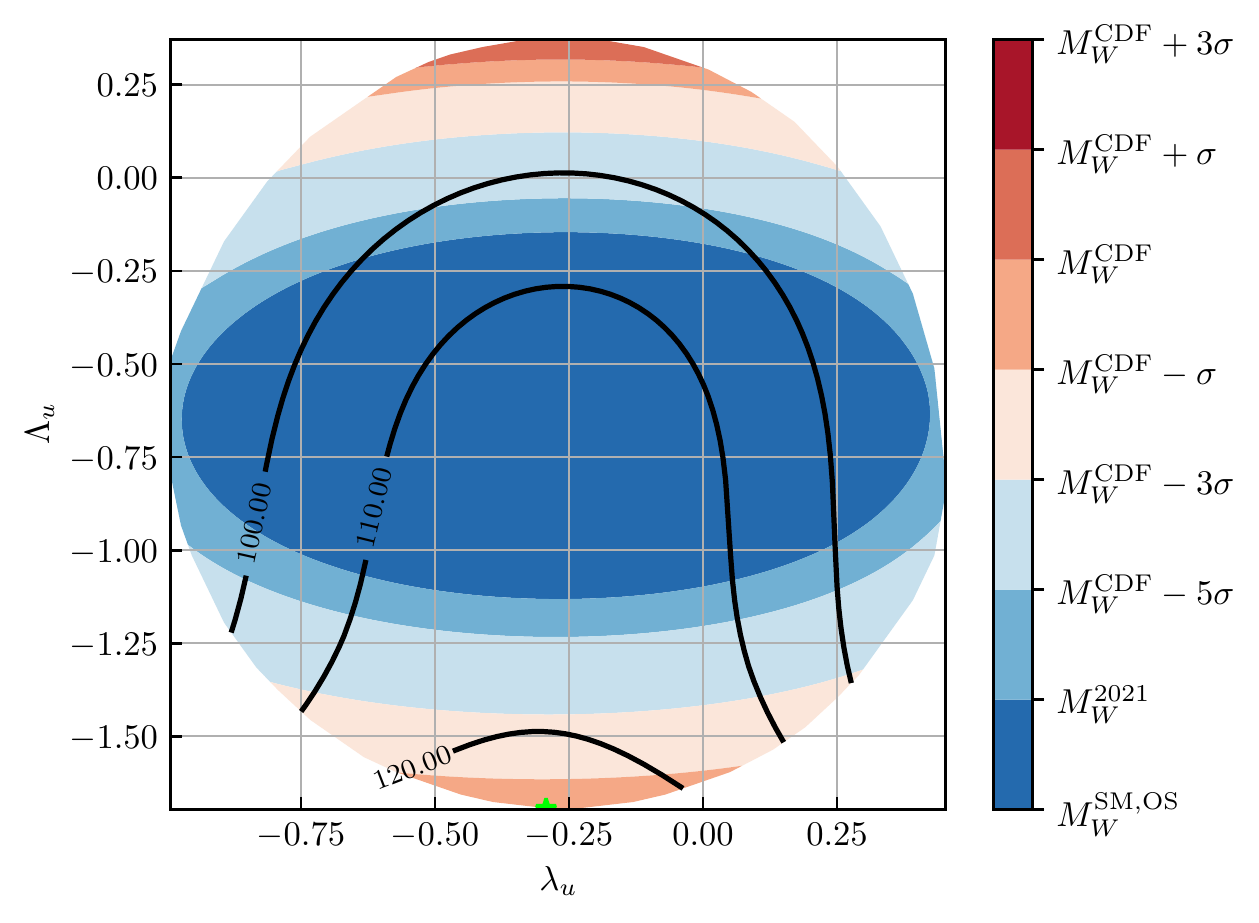}
 \caption{$\lambda_u$ vs.\ $\Lambda_u$}
 \label{fig:MRSSM-lamu_vs_Lamu}
 \end{subfigure}
 \begin{subfigure}{0.49\linewidth}
 \includegraphics[width=\textwidth]{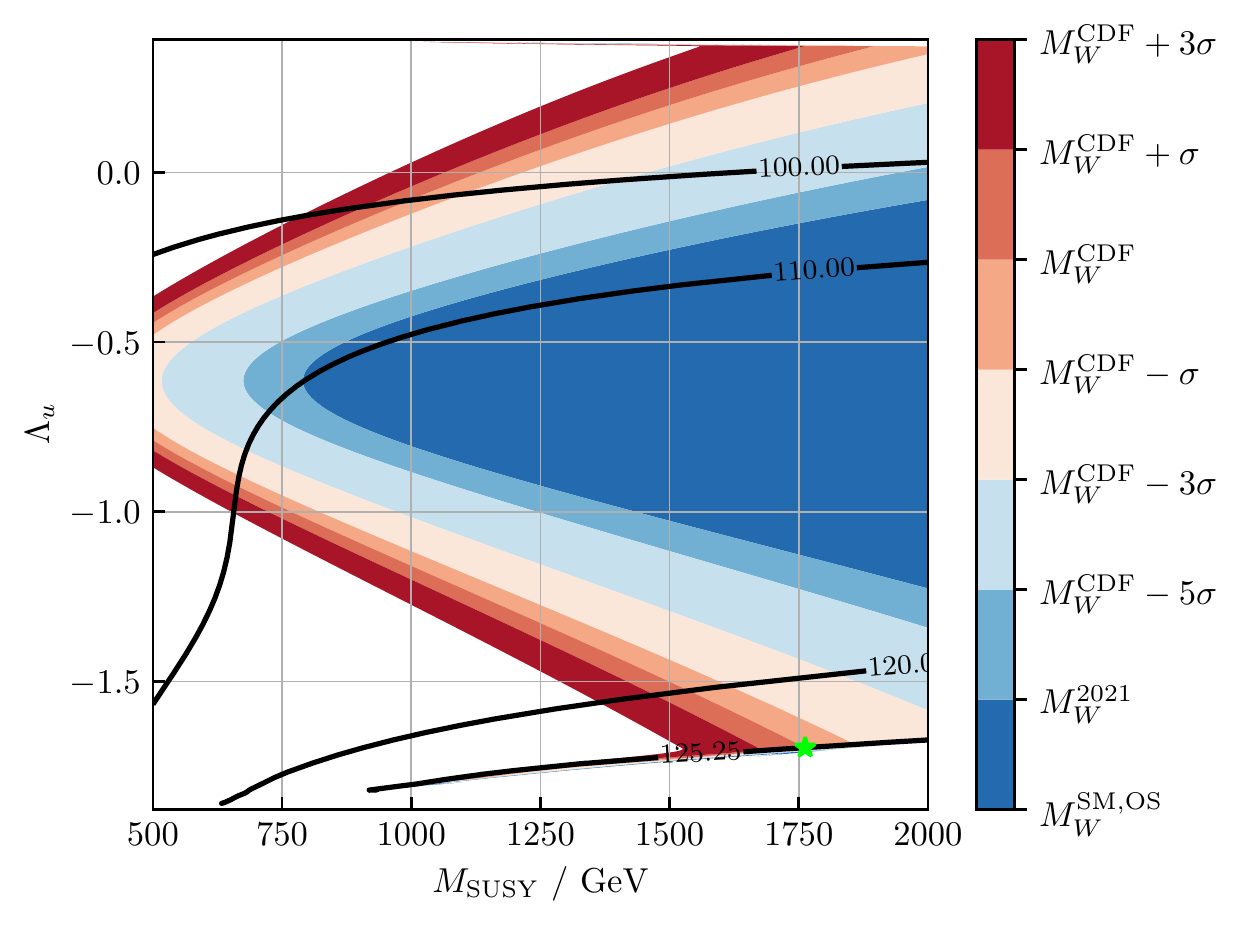}
 \caption{$\MSUSY$ vs.\ $\Lambda_u$}
 \label{fig:MRSSM-tanB_vs_Lamd}
 \end{subfigure}
 \caption{Prediction of the $W$ boson pole mass $M_W$ and contours for the
   Higgs boson pole mass $M_h$ in the \MRSSM\ around the
   point $\tan\beta \approx 24.7$, $\Lambda_d \approx -1.59$,
   $\Lambda_u=-1.70$, $\lambda_d\approx 2.18$, $\lambda_u\approx
   -0.293$ and $\MSUSY\approx 1763\GeV$ (marked with a $\star$). The
   color contours label landmark $M_W$ values in terms of the number
   of standard deviations from $M_W^\CDF$, or in the case of medium
   and dark blue bands, the 2021 world average and the \SM\ on-shell
   prediction, respectively. As such the value for $M_W$ varies a lot among
   the colored regions.}
 \label{fig:MRSSM_2d_plots}
\end{figure}

Moving away from the scenario \eqref{eq:MRSSM_scenario} we find that
both $M_h$ and $M_W^\CDF$ can be accommodated in a scenario with $\tan
\beta \approx 24.7$, 
$\Lambda_d \approx -1.59$, 
$\Lambda_u \approx -1.70$, 
$\lambda_d\approx 2.18$, 
$\lambda_u\approx -0.293$ 
and $\MSUSY\approx 1763\GeV$.
\footnote{The \FS input and the
  SLHA~\cite{Skands:2003cj,Allanach:2008qq} spectrum files for this
  point are attached to the \texttt{arXiv} submission of this
  publication.}  In \figref{fig:MRSSM_2d_plots} we show 2-dimensional
parameter scans around this point (marked with a green star symbol
$\star$) together with contours showing the value of the Higgs boson
pole mass $M_h$.  As shown in \figref{fig:MRSSM-lamu_vs_Lamu}, the
crucial parameters to obtain a correct $W$ boson pole mass are the
aforementioned top-Yukawa-like parameters $\lambda_u$ and
$\Lambda_u$. This is in agreement with previous findings in
Ref.~\cite{Diessner:2014ksa}.  The shape also suggests a quadratic
dependence of $M_W$ on the parameters, which fits with the analytic
expressions that have been obtained for simplified cases, see
Eq.~(4.16) of Ref.~\cite{Diessner:2014ksa}. The dependence of $M_W$ on
$\lambda_d$ and $\Lambda_d$ is very weak and therefore not shown.
Focusing now on $\Lambda_u$ and \MSUSY\ we can see that while the
\CDF\ $W$ mass measurement can be explained for almost any
\MSUSY\ value by adjusting $\Lambda_u$, combining this with the Higgs
mass measurement selects a rather narrow \MSUSY range, of about
$1.7$--$1.8\TeV$ (within $1\sigma$) from amongst the scenarios we
consider in this scan.  In both panels the white regions at the edges
are unphysical.  The color contours for $M_W$ are fixed by deviations
around the \CDF\ measurement, the 2021 world average and the
\SM\ on-shell prediction.  As such the colors do not represent even
shifts in the $W$ boson mass and care should be taken to avoid
confusion from this. Black contour lines show different values of the
\SM-like Higgs $M_h$, in particular these include the measured value
$M_h = 125.25\GeV$.\footnote{For these smaller $\MSUSY$ values we
  estimate the uncertainty of our fixed order one-loop calculation to
  be at most $\Delta M_h\approx 4\GeV$
  \cite{Athron:2016fuq}.\label{foo:Mh_uncert}}

\section{Conclusions}
\label{sec:conclusions}

In this short paper we have presented the new \FS\ calculation of
$M_W$ in a modified \MSbar/\DRbar-renormalization scheme.  Our new
calculation ensures decoupling with the precise \MSbar\ Standard Model
prediction being recovered in the limit where \BSM\ masses are much
larger than the Standard Model masses.  The Standard Model
contribution includes all currently known contributions, obtained
through the fit formula from the state-of-the-art \MSbar\ computation
in Ref.~\cite{Degrassi:2014sxa}. The \BSM\ contributions are
calculated at strict one-loop level avoiding the introduction of
spurious higher order terms.

We have briefly illustrated the calculation in the \MSSM, \SSM\ and the
\MRSSM.  In the \MRSSM\ we have also shown that this model can explain
the large deviation observed in the recent 2022 \CDF\ measurement of
the $W$ boson mass, which is the most precise measurement to date. We
believe that this updated calculation is very timely and will be very
useful for testing proposed explanations of this intriguing new
result, as well as a general constraint on Standard Model extensions.

\section*{Acknowledgements}

WK was supported by the National Science Centre (Poland) under the
research grant 2020/\allowbreak38/\allowbreak E/\allowbreak
ST2/\allowbreak00126.  DJ was supported by the Australian Government
Research Training Program (RTP) Scholarship, Monash Graduate
Excellence Scholarship (MGES), and the Deutscher Akademischer
Austauschdienst (DAAD) One-Year Research Grant. PA is supported by the
National Natural Science Foundation of China (NNSFC) Research Fund for
International Excellent Young Scientists grant 1215061046. MB and DS acknowledge support by DFG grant STO876/2.

\clearpage

\addcontentsline{toc}{section}{References}
\bibliography{bibliography}

\providecommand{\href}[2]{#2}\begingroup\raggedright\begin{thebibliography}{100}

\bibitem{Awramik:2003rn}
M.~Awramik, M.~Czakon, A.~Freitas and G.~Weiglein, \emph{{Precise prediction
  for the W boson mass in the standard model}},
  \href{http://dx.doi.org/10.1103/PhysRevD.69.053006}{\emph{Phys. Rev. D}
  \textbf{69} (2004) 053006},
  [\href{http://arxiv.org/abs/hep-ph/0311148}{hep-ph/0311148}].

\bibitem{Degrassi:2014sxa}
G.~Degrassi, P.~Gambino and P.~P. Giardino, \emph{{The $m_{\scriptscriptstyle
  W}-m_{\scriptscriptstyle Z}$ interdependence in the Standard Model: a new
  scrutiny}}, \href{http://dx.doi.org/10.1007/JHEP05(2015)154}{\emph{JHEP}
  \textbf{05} (2015) 154}, [\href{http://arxiv.org/abs/1411.7040}{1411.7040}].

\bibitem{Haller:2018nnx}
J.~Haller, A.~Hoecker, R.~Kogler, K.~M\"onig, T.~Peiffer and J.~Stelzer,
  \emph{{Update of the global electroweak fit and constraints on
  two-Higgs-doublet models}},
  \href{http://dx.doi.org/10.1140/epjc/s10052-018-6131-3}{\emph{Eur. Phys. J.
  C} \textbf{78} (2018) 675},
  [\href{http://arxiv.org/abs/1803.01853}{1803.01853}].

\bibitem{deBlas:2021wap}
J.~de~Blas, M.~Ciuchini, E.~Franco, A.~Goncalves, S.~Mishima, M.~Pierini
  et~al., \emph{{Global analysis of electroweak data in the Standard Model}},
  \href{http://arxiv.org/abs/2112.07274}{2112.07274}.

\bibitem{ALEPH:2006cdc}
{\scshape ALEPH} collaboration, S.~Schael et~al., \emph{{Measurement of the $W$
  boson mass and width in $e^{+} e^{-}$ collisions at LEP}},
  \href{http://dx.doi.org/10.1140/epjc/s2006-02576-8}{\emph{Eur. Phys. J. C}
  \textbf{47} (2006) 309--335},
  [\href{http://arxiv.org/abs/hep-ex/0605011}{hep-ex/0605011}].

\bibitem{DELPHI:2008avl}
{\scshape DELPHI} collaboration, J.~Abdallah et~al., \emph{{Measurement of the
  Mass and Width of the $W$ Boson in $e^{+} e^{-}$ Collisions at $\sqrt{s}$ =
  161-GeV - 209-GeV}},
  \href{http://dx.doi.org/10.1140/epjc/s10052-008-0585-7}{\emph{Eur. Phys. J.
  C} \textbf{55} (2008) 1--38},
  [\href{http://arxiv.org/abs/0803.2534}{0803.2534}].

\bibitem{L3:2005fft}
{\scshape L3} collaboration, P.~Achard et~al., \emph{{Measurement of the mass
  and the width of the $W$ boson at LEP}},
  \href{http://dx.doi.org/10.1140/epjc/s2005-02459-6}{\emph{Eur. Phys. J. C}
  \textbf{45} (2006) 569--587},
  [\href{http://arxiv.org/abs/hep-ex/0511049}{hep-ex/0511049}].

\bibitem{OPAL:2005rdt}
{\scshape OPAL} collaboration, G.~Abbiendi et~al., \emph{{Measurement of the
  mass and width of the $W$ boson}},
  \href{http://dx.doi.org/10.1140/epjc/s2005-02440-5}{\emph{Eur. Phys. J. C}
  \textbf{45} (2006) 307--335},
  [\href{http://arxiv.org/abs/hep-ex/0508060}{hep-ex/0508060}].

\bibitem{ALEPH:2013dgf}
{\scshape ALEPH, DELPHI, L3, OPAL, LEP Electroweak} collaboration, S.~Schael
  et~al., \emph{{Electroweak Measurements in Electron-Positron Collisions at
  W-Boson-Pair Energies at LEP}},
  \href{http://dx.doi.org/10.1016/j.physrep.2013.07.004}{\emph{Phys. Rept.}
  \textbf{532} (2013) 119--244},
  [\href{http://arxiv.org/abs/1302.3415}{1302.3415}].

\bibitem{D0:2012kms}
{\scshape D0} collaboration, V.~M. Abazov et~al., \emph{{Measurement of the W
  Boson Mass with the D0 Detector}},
  \href{http://dx.doi.org/10.1103/PhysRevLett.108.151804}{\emph{Phys. Rev.
  Lett.} \textbf{108} (2012) 151804},
  [\href{http://arxiv.org/abs/1203.0293}{1203.0293}].

\bibitem{CDF:2012gpf}
{\scshape CDF} collaboration, T.~Aaltonen et~al., \emph{{Precise measurement of
  the $W$-boson mass with the CDF II detector}},
  \href{http://dx.doi.org/10.1103/PhysRevLett.108.151803}{\emph{Phys. Rev.
  Lett.} \textbf{108} (2012) 151803},
  [\href{http://arxiv.org/abs/1203.0275}{1203.0275}].

\bibitem{CDF:2013dpa}
{\scshape CDF, D0} collaboration, T.~A. Aaltonen et~al., \emph{{Combination of
  CDF and D0 $W$-Boson Mass Measurements}},
  \href{http://dx.doi.org/10.1103/PhysRevD.88.052018}{\emph{Phys. Rev. D}
  \textbf{88} (2013) 052018},
  [\href{http://arxiv.org/abs/1307.7627}{1307.7627}].

\bibitem{ATLAS:2017rzl}
{\scshape ATLAS} collaboration, M.~Aaboud et~al., \emph{{Measurement of the
  $W$-boson mass in pp collisions at $\sqrt{s}=7$ TeV with the ATLAS
  detector}},
  \href{http://dx.doi.org/10.1140/epjc/s10052-017-5475-4}{\emph{Eur. Phys. J.
  C} \textbf{78} (2018) 110},
  [\href{http://arxiv.org/abs/1701.07240}{1701.07240}].

\bibitem{LHCb:2021bjt}
{\scshape LHCb} collaboration, R.~Aaij et~al., \emph{{Measurement of the W
  boson mass}}, \href{http://dx.doi.org/10.1007/JHEP01(2022)036}{\emph{JHEP}
  \textbf{01} (2022) 036},
  [\href{http://arxiv.org/abs/2109.01113}{2109.01113}].

\bibitem{ParticleDataGroup:2020ssz}
{\scshape Particle Data Group} collaboration, P.~A. Zyla et~al., \emph{{Review
  of Particle Physics}},
  \href{http://dx.doi.org/10.1093/ptep/ptaa104}{\emph{PTEP} \textbf{2020}
  (2020) 083C01}.

\bibitem{shortauthordoi:10.1126/science.abk1781}
T.~Aaltonen et~al., \emph{High-precision measurement of the <i>w</i> boson mass
  with the cdf ii detector},
  \href{http://dx.doi.org/10.1126/science.abk1781}{\emph{Science} \textbf{376}
  (2022) 170--176},
  [\href{http://arxiv.org/abs/https://www.science.org/doi/pdf/10.1126/science.abk1781}{https://www.science.org/doi/pdf/10.1126/science.abk1781}].

\bibitem{Lu:2022bgw}
C.-T. Lu, L.~Wu, Y.~Wu and B.~Zhu, \emph{{Electroweak Precision Fit and New
  Physics in light of $W$ Boson Mass}},
  \href{http://arxiv.org/abs/2204.03796}{2204.03796}.

\bibitem{Fan:2022dck}
Y.-Z. Fan, T.-P. Tang, Y.-L.~S. Tsai and L.~Wu, \emph{{Inert Higgs Dark Matter
  for New CDF W-boson Mass and Detection Prospects}},
  \href{http://arxiv.org/abs/2204.03693}{2204.03693}.

\bibitem{Athron:2022qpo}
P.~Athron, A.~Fowlie, C.-T. Lu, L.~Wu, Y.~Wu and B.~Zhu, \emph{{The $W$ boson
  Mass and Muon $g-2$: Hadronic Uncertainties or New Physics?}},
  \href{http://arxiv.org/abs/2204.03996}{2204.03996}.

\bibitem{Muong-2:2021ojo}
{\scshape Muon g-2} collaboration, B.~Abi et~al., \emph{{Measurement of the
  Positive Muon Anomalous Magnetic Moment to 0.46 ppm}},
  \href{http://dx.doi.org/10.1103/PhysRevLett.126.141801}{\emph{Phys. Rev.
  Lett.} \textbf{126} (2021) 141801},
  [\href{http://arxiv.org/abs/2104.03281}{2104.03281}].

\bibitem{Yuan:2022cpw}
G.-W. Yuan, L.~Zu, L.~Feng and Y.-F. Cai, \emph{{$W$-boson mass anomaly:
  probing the models of axion-like particle, dark photon and Chameleon dark
  energy}},  \href{http://arxiv.org/abs/2204.04183}{2204.04183}.

\bibitem{Strumia:2022qkt}
A.~Strumia, \emph{{Interpreting electroweak precision data including the
  $W$-mass CDF anomaly}},  \href{http://arxiv.org/abs/2204.04191}{2204.04191}.

\bibitem{Yang:2022gvz}
J.~M. Yang and Y.~Zhang, \emph{{Low energy SUSY confronted with new
  measurements of W-boson mass and muon g-2}},
  \href{http://arxiv.org/abs/2204.04202}{2204.04202}.

\bibitem{Du:2022pbp}
X.~K. Du, Z.~Li, F.~Wang and Y.~K. Zhang, \emph{{Explaining The Muon $g-2$
  Anomaly and New CDFII W-Boson Mass in the Framework of ExtraOrdinary Gauge
  Mediation}},  \href{http://arxiv.org/abs/2204.04286}{2204.04286}.

\bibitem{Zhu:2022tpr}
C.-R. Zhu, M.-Y. Cui, Z.-Q. Xia, Z.-H. Yu, X.~Huang, Q.~Yuan et~al., \emph{{GeV
  antiproton/gamma-ray excesses and the $W$-boson mass anomaly: three faces of
  $\sim 60-70$ GeV dark matter particle?}},
  \href{http://arxiv.org/abs/2204.03767}{2204.03767}.

\bibitem{deBlas:2022hdk}
J.~de~Blas, M.~Pierini, L.~Reina and L.~Silvestrini, \emph{{Impact of the
  recent measurements of the top-quark and W-boson masses on electroweak
  precision fits}},  \href{http://arxiv.org/abs/2204.04204}{2204.04204}.

\bibitem{Athron:2014yba}
P.~Athron, J.-h. Park, D.~St\"ockinger and A.~Voigt,
  \emph{{FlexibleSUSY\textemdash{}A spectrum generator generator for
  supersymmetric models}},
  \href{http://dx.doi.org/10.1016/j.cpc.2014.12.020}{\emph{Comput. Phys.
  Commun.} \textbf{190} (2015) 139--172},
  [\href{http://arxiv.org/abs/1406.2319}{1406.2319}].

\bibitem{Athron:2017fvs}
P.~Athron, M.~Bach, D.~Harries, T.~Kwasnitza, J.-h. Park, D.~St\"ockinger
  et~al., \emph{{FlexibleSUSY 2.0: Extensions to investigate the phenomenology
  of SUSY and non-SUSY models}},
  \href{http://dx.doi.org/10.1016/j.cpc.2018.04.016}{\emph{Comput. Phys.
  Commun.} \textbf{230} (2018) 145--217},
  [\href{http://arxiv.org/abs/1710.03760}{1710.03760}].

\bibitem{Athron:2021kve}
P.~Athron, A.~B\"uchner, D.~Harries, W.~Kotlarski, D.~St\"ockinger and
  A.~Voigt, \emph{{FlexibleDecay: An automated calculator of scalar decay
  widths}},  \href{http://arxiv.org/abs/2106.05038}{2106.05038}.

\bibitem{Diessner:2019ebm}
P.~Diessner and G.~Weiglein, \emph{{Precise prediction for the W boson mass in
  the MRSSM}}, \href{http://dx.doi.org/10.1007/JHEP07(2019)011}{\emph{JHEP}
  \textbf{07} (2019) 011},
  [\href{http://arxiv.org/abs/1904.03634}{1904.03634}].

\bibitem{Kribs:2007ac}
G.~D. Kribs, E.~Poppitz and N.~Weiner, \emph{{Flavor in supersymmetry with an
  extended R-symmetry}},
  \href{http://dx.doi.org/10.1103/PhysRevD.78.055010}{\emph{Phys. Rev. D}
  \textbf{78} (2008) 055010},
  [\href{http://arxiv.org/abs/0712.2039}{0712.2039}].

\bibitem{Heinemeyer:2013dia}
S.~Heinemeyer, W.~Hollik, G.~Weiglein and L.~Zeune, \emph{{Implications of LHC
  search results on the W boson mass prediction in the MSSM}},
  \href{http://dx.doi.org/10.1007/JHEP12(2013)084}{\emph{JHEP} \textbf{12}
  (2013) 084}, [\href{http://arxiv.org/abs/1311.1663}{1311.1663}].

\bibitem{Bagnaschi:2022qhb}
E.~Bagnaschi, M.~Chakraborti, S.~Heinemeyer, I.~Saha and G.~Weiglein,
  \emph{{Interdependence of the new ''MUON G-2'' Result and the $W$-Boson
  Mass}},  \href{http://arxiv.org/abs/2203.15710}{2203.15710}.

\bibitem{Sirlin:1980nh}
A.~Sirlin, \emph{{Radiative Corrections in the SU(2)-L x U(1) Theory: A Simple
  Renormalization Framework}},
  \href{http://dx.doi.org/10.1103/PhysRevD.22.971}{\emph{Phys. Rev. D}
  \textbf{22} (1980) 971--981}.

\bibitem{Marciano:1980pb}
W.~J. Marciano and A.~Sirlin, \emph{{Radiative Corrections to Neutrino Induced
  Neutral Current Phenomena in the SU(2)-L x U(1) Theory}},
  \href{http://dx.doi.org/10.1103/PhysRevD.22.2695}{\emph{Phys. Rev. D}
  \textbf{22} (1980) 2695}.

\bibitem{Sirlin:1983ys}
A.~Sirlin, \emph{{On the O(alpha**2) Corrections to tau (mu), m (W), m (Z) in
  the SU(2)-L x U(1) Theory}},
  \href{http://dx.doi.org/10.1103/PhysRevD.29.89}{\emph{Phys. Rev. D}
  \textbf{29} (1984) 89}.

\bibitem{Djouadi:1987gn}
A.~Djouadi and C.~Verzegnassi, \emph{{Virtual Very Heavy Top Effects in LEP /
  SLC Precision Measurements}},
  \href{http://dx.doi.org/10.1016/0370-2693(87)91206-8}{\emph{Phys. Lett. B}
  \textbf{195} (1987) 265--271}.

\bibitem{Djouadi:1987di}
A.~Djouadi, \emph{{O(alpha alpha-s) Vacuum Polarization Functions of the
  Standard Model Gauge Bosons}},
  \href{http://dx.doi.org/10.1007/BF02812964}{\emph{Nuovo Cim. A} \textbf{100}
  (1988) 357}.

\bibitem{Kniehl:1989yc}
B.~A. Kniehl, \emph{{Two Loop Corrections to the Vacuum Polarizations in
  Perturbative QCD}},
  \href{http://dx.doi.org/10.1016/0550-3213(90)90552-O}{\emph{Nucl. Phys. B}
  \textbf{347} (1990) 86--104}.

\bibitem{Consoli:1989fg}
M.~Consoli, W.~Hollik and F.~Jegerlehner, \emph{{The Effect of the Top Quark on
  the M(W)-M(Z) Interdependence and Possible Decoupling of Heavy Fermions from
  Low-Energy Physics}},
  \href{http://dx.doi.org/10.1016/0370-2693(89)91301-4}{\emph{Phys. Lett. B}
  \textbf{227} (1989) 167--170}.

\bibitem{Halzen:1990je}
F.~Halzen and B.~A. Kniehl, \emph{{$\Delta$ r beyond one loop}},
  \href{http://dx.doi.org/10.1016/0550-3213(91)90319-S}{\emph{Nucl. Phys. B}
  \textbf{353} (1991) 567--590}.

\bibitem{Kniehl:1991gu}
B.~A. Kniehl and A.~Sirlin, \emph{{Dispersion relations for vacuum polarization
  functions in electroweak physics}},
  \href{http://dx.doi.org/10.1016/0550-3213(92)90232-Z}{\emph{Nucl. Phys. B}
  \textbf{371} (1992) 141--148}.

\bibitem{Barbieri:1992nz}
R.~Barbieri, M.~Beccaria, P.~Ciafaloni, G.~Curci and A.~Vicere,
  \emph{{Radiative correction effects of a very heavy top}},
  \href{http://dx.doi.org/10.1016/0370-2693(92)91960-H}{\emph{Phys. Lett. B}
  \textbf{288} (1992) 95--98},
  [\href{http://arxiv.org/abs/hep-ph/9205238}{hep-ph/9205238}].

\bibitem{Djouadi:1993ss}
A.~Djouadi and P.~Gambino, \emph{{Electroweak gauge bosons selfenergies:
  Complete QCD corrections}},
  \href{http://dx.doi.org/10.1103/PhysRevD.49.3499}{\emph{Phys. Rev. D}
  \textbf{49} (1994) 3499--3511},
  [\href{http://arxiv.org/abs/hep-ph/9309298}{hep-ph/9309298}].

\bibitem{Fleischer:1993ub}
J.~Fleischer, O.~V. Tarasov and F.~Jegerlehner, \emph{{Two loop heavy top
  corrections to the rho parameter: A Simple formula valid for arbitrary Higgs
  mass}}, \href{http://dx.doi.org/10.1016/0370-2693(93)90810-5}{\emph{Phys.
  Lett. B} \textbf{319} (1993) 249--256}.

\bibitem{Degrassi:1996mg}
G.~Degrassi, P.~Gambino and A.~Vicini, \emph{{Two loop heavy top effects on the
  m(Z) - m(W) interdependence}},
  \href{http://dx.doi.org/10.1016/0370-2693(96)00720-4}{\emph{Phys. Lett. B}
  \textbf{383} (1996) 219--226},
  [\href{http://arxiv.org/abs/hep-ph/9603374}{hep-ph/9603374}].

\bibitem{Degrassi:1996ps}
G.~Degrassi, P.~Gambino and A.~Sirlin, \emph{{Precise calculation of M(W),
  sin**2 theta(W) (M(Z)), and sin**2 theta(eff)(lept)}},
  \href{http://dx.doi.org/10.1016/S0370-2693(96)01677-2}{\emph{Phys. Lett. B}
  \textbf{394} (1997) 188--194},
  [\href{http://arxiv.org/abs/hep-ph/9611363}{hep-ph/9611363}].

\bibitem{Freitas:2000gg}
A.~Freitas, W.~Hollik, W.~Walter and G.~Weiglein, \emph{{Complete fermionic two
  loop results for the M(W) - M(Z) interdependence}},
  \href{http://dx.doi.org/10.1016/S0370-2693(00)01263-6}{\emph{Phys. Lett. B}
  \textbf{495} (2000) 338--346},
  [\href{http://arxiv.org/abs/hep-ph/0007091}{hep-ph/0007091}].

\bibitem{Freitas:2002ja}
A.~Freitas, W.~Hollik, W.~Walter and G.~Weiglein, \emph{{Electroweak two loop
  corrections to the $M_W-M_Z$ mass correlation in the standard model}},
  \href{http://dx.doi.org/10.1016/S0550-3213(02)00243-2}{\emph{Nucl. Phys. B}
  \textbf{632} (2002) 189--218},
  [\href{http://arxiv.org/abs/hep-ph/0202131}{hep-ph/0202131}].

\bibitem{Awramik:2002wn}
M.~Awramik and M.~Czakon, \emph{{Complete two loop bosonic contributions to the
  muon lifetime in the standard model}},
  \href{http://dx.doi.org/10.1103/PhysRevLett.89.241801}{\emph{Phys. Rev.
  Lett.} \textbf{89} (2002) 241801},
  [\href{http://arxiv.org/abs/hep-ph/0208113}{hep-ph/0208113}].

\bibitem{Awramik:2003ee}
M.~Awramik and M.~Czakon, \emph{{Complete two loop electroweak contributions to
  the muon lifetime in the standard model}},
  \href{http://dx.doi.org/10.1016/j.physletb.2003.06.007}{\emph{Phys. Lett. B}
  \textbf{568} (2003) 48--54},
  [\href{http://arxiv.org/abs/hep-ph/0305248}{hep-ph/0305248}].

\bibitem{Onishchenko:2002ve}
A.~Onishchenko and O.~Veretin, \emph{{Two loop bosonic electroweak corrections
  to the muon lifetime and M(Z) - M(W) interdependence}},
  \href{http://dx.doi.org/10.1016/S0370-2693(02)03004-6}{\emph{Phys. Lett. B}
  \textbf{551} (2003) 111--114},
  [\href{http://arxiv.org/abs/hep-ph/0209010}{hep-ph/0209010}].

\bibitem{Awramik:2002vu}
M.~Awramik, M.~Czakon, A.~Onishchenko and O.~Veretin, \emph{{Bosonic
  corrections to Delta r at the two loop level}},
  \href{http://dx.doi.org/10.1103/PhysRevD.68.053004}{\emph{Phys. Rev. D}
  \textbf{68} (2003) 053004},
  [\href{http://arxiv.org/abs/hep-ph/0209084}{hep-ph/0209084}].

\bibitem{Avdeev:1994db}
L.~Avdeev, J.~Fleischer, S.~Mikhailov and O.~Tarasov, \emph{{$0(\alpha
  \alpha_s^2)$ correction to the electroweak $\rho$ parameter}},
  \href{http://dx.doi.org/10.1016/0370-2693(94)90573-8}{\emph{Phys. Lett. B}
  \textbf{336} (1994) 560--566},
  [\href{http://arxiv.org/abs/hep-ph/9406363}{hep-ph/9406363}].

\bibitem{Chetyrkin:1995ix}
K.~G. Chetyrkin, J.~H. Kuhn and M.~Steinhauser, \emph{{Corrections of order
  ${\cal O}(G_F M_t^2 \alpha_s^2)$ to the $\rho$ parameter}},
  \href{http://dx.doi.org/10.1016/0370-2693(95)00380-4}{\emph{Phys. Lett. B}
  \textbf{351} (1995) 331--338},
  [\href{http://arxiv.org/abs/hep-ph/9502291}{hep-ph/9502291}].

\bibitem{Chetyrkin:1995js}
K.~G. Chetyrkin, J.~H. Kuhn and M.~Steinhauser, \emph{{QCD corrections from top
  quark to relations between electroweak parameters to order alpha-s**2}},
  \href{http://dx.doi.org/10.1103/PhysRevLett.75.3394}{\emph{Phys. Rev. Lett.}
  \textbf{75} (1995) 3394--3397},
  [\href{http://arxiv.org/abs/hep-ph/9504413}{hep-ph/9504413}].

\bibitem{Chetyrkin:1996cf}
K.~G. Chetyrkin, J.~H. Kuhn and M.~Steinhauser, \emph{{Three loop polarization
  function and O (alpha-s**2) corrections to the production of heavy quarks}},
  \href{http://dx.doi.org/10.1016/S0550-3213(96)00534-2}{\emph{Nucl. Phys. B}
  \textbf{482} (1996) 213--240},
  [\href{http://arxiv.org/abs/hep-ph/9606230}{hep-ph/9606230}].

\bibitem{Faisst:2003px}
M.~Faisst, J.~H. Kuhn, T.~Seidensticker and O.~Veretin, \emph{{Three loop top
  quark contributions to the rho parameter}},
  \href{http://dx.doi.org/10.1016/S0550-3213(03)00450-4}{\emph{Nucl. Phys. B}
  \textbf{665} (2003) 649--662},
  [\href{http://arxiv.org/abs/hep-ph/0302275}{hep-ph/0302275}].

\bibitem{vanderBij:2000cg}
J.~J. van~der Bij, K.~G. Chetyrkin, M.~Faisst, G.~Jikia and T.~Seidensticker,
  \emph{{Three loop leading top mass contributions to the rho parameter}},
  \href{http://dx.doi.org/10.1016/S0370-2693(01)00002-8}{\emph{Phys. Lett. B}
  \textbf{498} (2001) 156--162},
  [\href{http://arxiv.org/abs/hep-ph/0011373}{hep-ph/0011373}].

\bibitem{Boughezal:2004ef}
R.~Boughezal, J.~B. Tausk and J.~J. van~der Bij, \emph{{Three-loop electroweak
  correction to the Rho parameter in the large Higgs mass limit}},
  \href{http://dx.doi.org/10.1016/j.nuclphysb.2005.02.020}{\emph{Nucl. Phys. B}
  \textbf{713} (2005) 278--290},
  [\href{http://arxiv.org/abs/hep-ph/0410216}{hep-ph/0410216}].

\bibitem{Boughezal:2006xk}
R.~Boughezal and M.~Czakon, \emph{{Single scale tadpoles and O(G(F m(t)**2
  alpha(s)**3)) corrections to the rho parameter}},
  \href{http://dx.doi.org/10.1016/j.nuclphysb.2006.08.007}{\emph{Nucl. Phys. B}
  \textbf{755} (2006) 221--238},
  [\href{http://arxiv.org/abs/hep-ph/0606232}{hep-ph/0606232}].

\bibitem{Chetyrkin:2006bj}
K.~G. Chetyrkin, M.~Faisst, J.~H. Kuhn, P.~Maierhofer and C.~Sturm,
  \emph{{Four-Loop QCD Corrections to the Rho Parameter}},
  \href{http://dx.doi.org/10.1103/PhysRevLett.97.102003}{\emph{Phys. Rev.
  Lett.} \textbf{97} (2006) 102003},
  [\href{http://arxiv.org/abs/hep-ph/0605201}{hep-ph/0605201}].

\bibitem{Schroder:2005db}
Y.~Schroder and M.~Steinhauser, \emph{{Four-loop singlet contribution to the
  rho parameter}},
  \href{http://dx.doi.org/10.1016/j.physletb.2005.06.085}{\emph{Phys. Lett. B}
  \textbf{622} (2005) 124--130},
  [\href{http://arxiv.org/abs/hep-ph/0504055}{hep-ph/0504055}].

\bibitem{Garcia:1993sb}
D.~Garcia and J.~Sola, \emph{{Full one loop supersymmetric quantum effects on
  M(W)}}, \href{http://dx.doi.org/10.1142/S021773239400023X}{\emph{Mod. Phys.
  Lett. A} \textbf{9} (1994) 211--224}.

\bibitem{Chankowski:1993eu}
P.~H. Chankowski, A.~Dabelstein, W.~Hollik, W.~M. Mosle, S.~Pokorski and
  J.~Rosiek, \emph{{Delta R in the MSSM}},
  \href{http://dx.doi.org/10.1016/0550-3213(94)90539-8}{\emph{Nucl. Phys. B}
  \textbf{417} (1994) 101--129}.

\bibitem{Pierce:1996zz}
D.~M. Pierce, J.~A. Bagger, K.~T. Matchev and R.-j. Zhang, \emph{{Precision
  corrections in the minimal supersymmetric standard model}},
  \href{http://dx.doi.org/10.1016/S0550-3213(96)00683-9}{\emph{Nucl. Phys.}
  \textbf{B491} (1997) 3--67},
  [\href{http://arxiv.org/abs/hep-ph/9606211}{hep-ph/9606211}].

\bibitem{Barbieri:1983wy}
R.~Barbieri and L.~Maiani, \emph{{Renormalization of the Electroweak rho
  Parameter from Supersymmetric Particles}},
  \href{http://dx.doi.org/10.1016/0550-3213(83)90311-5}{\emph{Nucl. Phys. B}
  \textbf{224} (1983) 32}.

\bibitem{Lim:1983re}
C.~S. Lim, T.~Inami and N.~Sakai, \emph{{The $\rho$ Parameter in Supersymmetric
  Models}}, \href{http://dx.doi.org/10.1103/PhysRevD.29.1488}{\emph{Phys. Rev.
  D} \textbf{29} (1984) 1488}.

\bibitem{Eliasson:1984yu}
E.~Eliasson, \emph{{Radiative Corrections to Electroweak Interactions in
  Supergravity {GUTs}}},
  \href{http://dx.doi.org/10.1016/0370-2693(84)90593-8}{\emph{Phys. Lett. B}
  \textbf{147} (1984) 65--72}.

\bibitem{Hioki:1985wz}
Z.~Hioki, \emph{{One Loop Effects of Heavy Scalar Quarks in Supersymmetric
  Electroweak Theory}},
  \href{http://dx.doi.org/10.1143/PTP.73.1283}{\emph{Prog. Theor. Phys.}
  \textbf{73} (1985) 1283}.

\bibitem{Grifols:1984xs}
J.~A. Grifols and J.~Sola, \emph{{One Loop Renormalization of the Electroweak
  Parameters in $N=1$ Supersymmetry}},
  \href{http://dx.doi.org/10.1016/0550-3213(85)90519-X}{\emph{Nucl. Phys. B}
  \textbf{253} (1985) 47}.

\bibitem{Barbieri:1989dc}
R.~Barbieri, M.~Frigeni, F.~Giuliani and H.~E. Haber, \emph{{Precision
  Measurements in Electroweak Physics and Supersymmetry}},
  \href{http://dx.doi.org/10.1016/0550-3213(90)90181-C}{\emph{Nucl. Phys. B}
  \textbf{341} (1990) 309--321}.

\bibitem{Gosdzinsky:1990sk}
P.~Gosdzinsky and J.~Sola, \emph{{A Numerical analysis of the full one loop
  renormalization of the weak gauge boson masses from supersymmetry}},
  \href{http://dx.doi.org/10.1142/S0217732391002098}{\emph{Mod. Phys. Lett. A}
  \textbf{6} (1991) 1943--1952}.

\bibitem{Drees:1990dx}
M.~Drees and K.~Hagiwara, \emph{{Supersymmetric Contribution to the Electroweak
  $\rho$ Parameter}},
  \href{http://dx.doi.org/10.1103/PhysRevD.42.1709}{\emph{Phys. Rev. D}
  \textbf{42} (1990) 1709--1725}.

\bibitem{Drees:1991zk}
M.~Drees, K.~Hagiwara and A.~Yamada, \emph{{Process independent radiative
  corrections in the minimal supersymmetric standard model}},
  \href{http://dx.doi.org/10.1103/PhysRevD.45.1725}{\emph{Phys. Rev. D}
  \textbf{45} (1992) 1725--1743}.

\bibitem{Djouadi:1996pa}
A.~Djouadi, P.~Gambino, S.~Heinemeyer, W.~Hollik, C.~Junger and G.~Weiglein,
  \emph{{Supersymmetric contributions to electroweak precision observables: QCD
  corrections}},
  \href{http://dx.doi.org/10.1103/PhysRevLett.78.3626}{\emph{Phys. Rev. Lett.}
  \textbf{78} (1997) 3626--3629},
  [\href{http://arxiv.org/abs/hep-ph/9612363}{hep-ph/9612363}].

\bibitem{Djouadi:1998sq}
A.~Djouadi, P.~Gambino, S.~Heinemeyer, W.~Hollik, C.~Junger and G.~Weiglein,
  \emph{{Leading QCD corrections to scalar quark contributions to electroweak
  precision observables}},
  \href{http://dx.doi.org/10.1103/PhysRevD.57.4179}{\emph{Phys. Rev. D}
  \textbf{57} (1998) 4179--4196},
  [\href{http://arxiv.org/abs/hep-ph/9710438}{hep-ph/9710438}].

\bibitem{Heinemeyer:2002jq}
S.~Heinemeyer and G.~Weiglein, \emph{{Leading electroweak two loop corrections
  to precision observables in the MSSM}},
  \href{http://dx.doi.org/10.1088/1126-6708/2002/10/072}{\emph{JHEP}
  \textbf{10} (2002) 072},
  [\href{http://arxiv.org/abs/hep-ph/0209305}{hep-ph/0209305}].

\bibitem{Heinemeyer:2004gx}
S.~Heinemeyer, W.~Hollik and G.~Weiglein, \emph{{Electroweak precision
  observables in the minimal supersymmetric standard model}},
  \href{http://dx.doi.org/10.1016/j.physrep.2005.12.002}{\emph{Phys. Rept.}
  \textbf{425} (2006) 265--368},
  [\href{http://arxiv.org/abs/hep-ph/0412214}{hep-ph/0412214}].

\bibitem{Heinemeyer:2006px}
S.~Heinemeyer, W.~Hollik, D.~Stockinger, A.~M. Weber and G.~Weiglein,
  \emph{{Precise prediction for M(W) in the MSSM}},
  \href{http://dx.doi.org/10.1088/1126-6708/2006/08/052}{\emph{JHEP}
  \textbf{08} (2006) 052},
  [\href{http://arxiv.org/abs/hep-ph/0604147}{hep-ph/0604147}].

\bibitem{Lopez-Val:2012uou}
D.~Lopez-Val and J.~Sola, \emph{{Delta r in the Two-Higgs-Doublet Model at full
  one loop level -- and beyond}},
  \href{http://dx.doi.org/10.1140/epjc/s10052-013-2393-y}{\emph{Eur. Phys. J.
  C} \textbf{73} (2013) 2393},
  [\href{http://arxiv.org/abs/1211.0311}{1211.0311}].

\bibitem{Hessenberger:2016atw}
S.~Hessenberger and W.~Hollik, \emph{{Two-loop corrections to the $\rho$
  parameter in Two-Higgs-Doublet Models}},
  \href{http://dx.doi.org/10.1140/epjc/s10052-017-4734-8}{\emph{Eur. Phys. J.
  C} \textbf{77} (2017) 178},
  [\href{http://arxiv.org/abs/1607.04610}{1607.04610}].

\bibitem{Lopez-Val:2014jva}
D.~L\'opez-Val and T.~Robens, \emph{{\ensuremath{\Delta}r and the W-boson mass
  in the singlet extension of the standard model}},
  \href{http://dx.doi.org/10.1103/PhysRevD.90.114018}{\emph{Phys. Rev. D}
  \textbf{90} (2014) 114018},
  [\href{http://arxiv.org/abs/1406.1043}{1406.1043}].

\bibitem{Domingo:2011uf}
F.~Domingo and T.~Lenz, \emph{{W mass and Leptonic Z-decays in the NMSSM}},
  \href{http://dx.doi.org/10.1007/JHEP07(2011)101}{\emph{JHEP} \textbf{07}
  (2011) 101}, [\href{http://arxiv.org/abs/1101.4758}{1101.4758}].

\bibitem{Allanach:2013kza}
B.~C. Allanach, P.~Athron, L.~C. Tunstall, A.~Voigt and A.~G. Williams,
  \emph{{Next-to-Minimal SOFTSUSY}},
  \href{http://dx.doi.org/10.1016/j.cpc.2014.04.015}{\emph{Comput. Phys.
  Commun.} \textbf{185} (2014) 2322--2339},
  [\href{http://arxiv.org/abs/1311.7659}{1311.7659}].

\bibitem{Stal:2015zca}
O.~St\r{a}l, G.~Weiglein and L.~Zeune, \emph{{Improved prediction for the mass
  of the W boson in the NMSSM}},
  \href{http://dx.doi.org/10.1007/JHEP09(2015)158}{\emph{JHEP} \textbf{09}
  (2015) 158}, [\href{http://arxiv.org/abs/1506.07465}{1506.07465}].

\bibitem{Diessner:2014ksa}
P.~Die\ss{}ner, J.~Kalinowski, W.~Kotlarski and D.~St\"ockinger, \emph{{Higgs
  boson mass and electroweak observables in the MRSSM}},
  \href{http://dx.doi.org/10.1007/JHEP12(2014)124}{\emph{JHEP} \textbf{12}
  (2014) 124}, [\href{http://arxiv.org/abs/1410.4791}{1410.4791}].

\bibitem{Staub:2009bi}
F.~Staub, \emph{{From Superpotential to Model Files for FeynArts and
  CalcHep/CompHep}},
  \href{http://dx.doi.org/10.1016/j.cpc.2010.01.011}{\emph{Comput. Phys.
  Commun.} \textbf{181} (2010) 1077--1086},
  [\href{http://arxiv.org/abs/0909.2863}{0909.2863}].

\bibitem{Staub:2010jh}
F.~Staub, \emph{{Automatic Calculation of supersymmetric Renormalization Group
  Equations and Self Energies}},
  \href{http://dx.doi.org/10.1016/j.cpc.2010.11.030}{\emph{Comput. Phys.
  Commun.} \textbf{182} (2011) 808--833},
  [\href{http://arxiv.org/abs/1002.0840}{1002.0840}].

\bibitem{Staub:2012pb}
F.~Staub, \emph{{SARAH 3.2: Dirac Gauginos, UFO output, and more}},
  \href{http://dx.doi.org/10.1016/j.cpc.2013.02.019}{\emph{Comput. Phys.
  Commun.} \textbf{184} (2013) 1792--1809},
  [\href{http://arxiv.org/abs/1207.0906}{1207.0906}].

\bibitem{Staub:2013tta}
F.~Staub, \emph{{SARAH 4 : A tool for (not only SUSY) model builders}},
  \href{http://dx.doi.org/10.1016/j.cpc.2014.02.018}{\emph{Comput. Phys.
  Commun.} \textbf{185} (2014) 1773--1790},
  [\href{http://arxiv.org/abs/1309.7223}{1309.7223}].

\bibitem{Porod:2011nf}
W.~Porod and F.~Staub, \emph{{SPheno 3.1: Extensions including flavour,
  CP-phases and models beyond the MSSM}},
  \href{http://dx.doi.org/10.1016/j.cpc.2012.05.021}{\emph{Comput. Phys.
  Commun.} \textbf{183} (2012) 2458--2469},
  [\href{http://arxiv.org/abs/1104.1573}{1104.1573}].

\bibitem{Degrassi:1990tu}
G.~Degrassi, S.~Fanchiotti and A.~Sirlin, \emph{{Relations Between the On-shell
  and MS Frameworks and the $M_W$-$M_Z$ Interdependence}},
  \href{http://dx.doi.org/10.1016/0550-3213(91)90081-8}{\emph{Nucl. Phys.}
  \textbf{B351} (1991) 49--69}.

\bibitem{Allanach:2001kg}
B.~C. Allanach, \emph{{SOFTSUSY: a program for calculating supersymmetric
  spectra}},
  \href{http://dx.doi.org/10.1016/S0010-4655(01)00460-X}{\emph{Comput. Phys.
  Commun.} \textbf{143} (2002) 305--331},
  [\href{http://arxiv.org/abs/hep-ph/0104145}{hep-ph/0104145}].

\bibitem{Kwasnitza:2020wli}
T.~Kwasnitza, D.~St\"ockinger and A.~Voigt, \emph{{Improved MSSM Higgs mass
  calculation using the 3-loop FlexibleEFTHiggs approach including
  $x_{t}$-resummation}},
  \href{http://dx.doi.org/10.1007/JHEP07(2020)197}{\emph{JHEP} \textbf{07}
  (2020) 197}, [\href{http://arxiv.org/abs/2003.04639}{2003.04639}].

\bibitem{Degrassi:2001yf}
G.~Degrassi, P.~Slavich and F.~Zwirner, \emph{{On the neutral Higgs boson
  masses in the MSSM for arbitrary stop mixing}},
  \href{http://dx.doi.org/10.1016/S0550-3213(01)00343-1}{\emph{Nucl. Phys. B}
  \textbf{611} (2001) 403--422},
  [\href{http://arxiv.org/abs/hep-ph/0105096}{hep-ph/0105096}].

\bibitem{Brignole:2001jy}
A.~Brignole, G.~Degrassi, P.~Slavich and F.~Zwirner, \emph{{On the
  O(alpha(t)**2) two loop corrections to the neutral Higgs boson masses in the
  MSSM}}, \href{http://dx.doi.org/10.1016/S0550-3213(02)00184-0}{\emph{Nucl.
  Phys. B} \textbf{631} (2002) 195--218},
  [\href{http://arxiv.org/abs/hep-ph/0112177}{hep-ph/0112177}].

\bibitem{Dedes:2002dy}
A.~Dedes and P.~Slavich, \emph{{Two loop corrections to radiative electroweak
  symmetry breaking in the MSSM}},
  \href{http://dx.doi.org/10.1016/S0550-3213(03)00173-1}{\emph{Nucl. Phys. B}
  \textbf{657} (2003) 333--354},
  [\href{http://arxiv.org/abs/hep-ph/0212132}{hep-ph/0212132}].

\bibitem{Brignole:2002bz}
A.~Brignole, G.~Degrassi, P.~Slavich and F.~Zwirner, \emph{{On the two loop
  sbottom corrections to the neutral Higgs boson masses in the MSSM}},
  \href{http://dx.doi.org/10.1016/S0550-3213(02)00748-4}{\emph{Nucl. Phys. B}
  \textbf{643} (2002) 79--92},
  [\href{http://arxiv.org/abs/hep-ph/0206101}{hep-ph/0206101}].

\bibitem{Dedes:2003km}
A.~Dedes, G.~Degrassi and P.~Slavich, \emph{{On the two loop Yukawa corrections
  to the MSSM Higgs boson masses at large tan beta}},
  \href{http://dx.doi.org/10.1016/j.nuclphysb.2003.08.033}{\emph{Nucl. Phys. B}
  \textbf{672} (2003) 144--162},
  [\href{http://arxiv.org/abs/hep-ph/0305127}{hep-ph/0305127}].

\bibitem{Harlander:2008ju}
R.~V. Harlander, P.~Kant, L.~Mihaila and M.~Steinhauser, \emph{{Higgs boson
  mass in supersymmetry to three loops}},
  \href{http://dx.doi.org/10.1103/PhysRevLett.101.039901}{\emph{Phys. Rev.
  Lett.} \textbf{100} (2008) 191602},
  [\href{http://arxiv.org/abs/0803.0672}{0803.0672}].

\bibitem{Kant:2010tf}
P.~Kant, R.~V. Harlander, L.~Mihaila and M.~Steinhauser, \emph{{Light MSSM
  Higgs boson mass to three-loop accuracy}},
  \href{http://dx.doi.org/10.1007/JHEP08(2010)104}{\emph{JHEP} \textbf{08}
  (2010) 104}, [\href{http://arxiv.org/abs/1005.5709}{1005.5709}].

\bibitem{Kunz:2014gya}
D.~Kunz, L.~Mihaila and N.~Zerf, \emph{{$\mathcal O(\alpha_S^2)$ corrections to
  the running top-Yukawa coupling and the mass of the lightest Higgs boson in
  the MSSM}}, \href{http://dx.doi.org/10.1007/JHEP12(2014)136}{\emph{JHEP}
  \textbf{12} (2014) 136}, [\href{http://arxiv.org/abs/1409.2297}{1409.2297}].

\bibitem{Harlander:2017kuc}
R.~V. Harlander, J.~Klappert and A.~Voigt, \emph{{Higgs mass prediction in the
  MSSM at three-loop level in a pure $\overline{{\text {DR}}}$ context}},
  \href{http://dx.doi.org/10.1140/epjc/s10052-017-5368-6}{\emph{Eur. Phys. J.
  C} \textbf{77} (2017) 814},
  [\href{http://arxiv.org/abs/1708.05720}{1708.05720}].

\bibitem{Athron:2016fuq}
P.~Athron, J.-h. Park, T.~Steudtner, D.~St\"ockinger and A.~Voigt,
  \emph{{Precise Higgs mass calculations in (non-)minimal supersymmetry at both
  high and low scales}},
  \href{http://dx.doi.org/10.1007/JHEP01(2017)079}{\emph{JHEP} \textbf{01}
  (2017) 079}, [\href{http://arxiv.org/abs/1609.00371}{1609.00371}].

\bibitem{Skands:2003cj}
P.~Z. Skands et~al., \emph{{SUSY Les Houches accord: Interfacing SUSY spectrum
  calculators, decay packages, and event generators}},
  \href{http://dx.doi.org/10.1088/1126-6708/2004/07/036}{\emph{JHEP}
  \textbf{07} (2004) 036},
  [\href{http://arxiv.org/abs/hep-ph/0311123}{hep-ph/0311123}].

\bibitem{Allanach:2008qq}
B.~C. Allanach et~al., \emph{{SUSY Les Houches Accord 2}},
  \href{http://dx.doi.org/10.1016/j.cpc.2008.08.004}{\emph{Comput. Phys.
  Commun.} \textbf{180} (2009) 8--25},
  [\href{http://arxiv.org/abs/0801.0045}{0801.0045}].

\end{thebibliography}\endgroup
\bibliographystyle{JHEP}

\end{document}